\documentclass[12pt,nofootinbib,eqsecnum,longbibliography,prd]{revtex4-2}

\usepackage{amsmath,amssymb}
\usepackage[american]{babel}
\usepackage[export]{adjustbox}
\usepackage{microtype}
\usepackage{color}
\usepackage{graphicx}
\usepackage{hyperref}
\hypersetup{colorlinks, citecolor=blue, linkcolor=black, urlcolor=blue}
\usepackage{bm}
\usepackage{tikz-feynman} 
\usepackage{anyfontsize}

\newcommand{\ket}[1]{|#1\rangle}

\renewcommand{\v}[1]{\mathbf{#1}}

\newcommand{\sign}{\mathrm{sign}}
\newcommand{\newvar}{b}

\definecolor{purple}{rgb}{0.4,0.0,0.4}

\begin{document}

\title{Quantum Electrodynamics Mediated by a Photon with Generalized (Continuous) Spin}

\author{Philip Schuster}
\email{schuster@slac.stanford.edu}
\affiliation{SLAC National Accelerator Laboratory, 2575 Sand Hill Road, Menlo Park, CA 94025, USA}
\author{Natalia Toro}
\email{ntoro@slac.stanford.edu}
\affiliation{SLAC National Accelerator Laboratory, 2575 Sand Hill Road, Menlo Park, CA 94025, USA}

\date{\today}

\begin{abstract}
We present rules for computing scattering amplitudes of charged scalar matter and photons, where the photon has non-zero spin Casimir $\rho$, and is therefore a continuous spin particle (CSP). The amplitudes reduce to familiar scalar QED when $\rho\rightarrow 0$. As a concrete example, we compute the pair annihilation and Compton scattering amplitudes in this theory and comment on their physical properties, including unitarity and scaling behavior at small and large $\rho$.
\end{abstract}

\maketitle
\newpage
\tableofcontents



\newpage
\section{Introduction}
\label{Introduction}

Understanding the most general properties of long-range physics compatible with Lorentz symmetry and quantum mechanics is both fascinating and important. It is also an unresolved question in certain simple and surprisingly fundamental respects. 

The most general type of massless state in 3+1 dimensions is described by an irreducible representation of the ISO(2) Little Group\cite{Wigner:1939cj,WeinbergQFT}. Beyond the mass $k^2=m^2=0$, the other invariant quantum number of a massless particle is the square of the Pauli-Lubanski vector (the spin Casimir) $W^2 = - \rho^2$, where we call $\rho$ the spin scale. 
For the case of bosons, a massless state $\ket{k,h}$ is labeled by its momentum $k$ and ``helicity'' $h$, the eigenvalue of rotations about the direction of the spatial momentum.  When $\rho=0$, these polarizations are boost invariant, can be described singly, and arise in familiar gauge theories such as QED.  When $\rho \neq 0$, an infinite tower of polarization modes at integer $h$ mix with one another under boosts (much like the $2s+1$ polarization states of a massive spin-$s$ particle)
\footnote{In 2+1 dimensions, the elementary state has one polarization, called a {\it panyon}, wich is also a certain massless limit of an anyon\cite{Schuster:2014xja}.}. Until recently, not much was known about interactions in the more general case $\rho\neq 0$. 
%
%

In \cite{Schuster:2023xqa}, we and Zhou presented a classical formalism appropriate for computing interactions of scalar matter with massless gauge fields \cite{Schuster:2014hca} describing states with generalized $\rho\neq 0$ (i.e. continuous) spin, a.k.a.~continuous spin particles (CSPs).  
Though many questions remain about the space-time properties of this formalism, we have used it to derive rules for computing quantum scattering amplitudes in a generalization of scalar QED with $\rho\neq 0$. The amplitudes reduce to familiar scalar QED when $\rho\rightarrow 0$, with all but the $\pm 1$ helicity modes of the CSP photon decoupling from the matter in this limit. 


Our generalization of scalar QED to $\rho\neq 0$ is formulated in the first-quantized path integral formalism for scalar matter, pioneered by Feynman in the case of quantum electrodynamics \cite{Feynman:1950ir}, and appropriately combined with the field theory formalism of \cite{Schuster:2023xqa}. For the matter, the transition elements will be computed from a $0+1$-dimensional quantum field theory with underlying Minkowski target space (a scalar worldline) and interactions encoded by vertex operators corresponding to external CSP photon states. This will be very familiar to string theorists and those familiar with the first quantized methods for computing gauge theory amplitudes. For off-shell CSP photons, we will use the the appropriate propagator from \cite{Schuster:2023xqa} in a convenient gauge. We build on the string-inspired computational approach developed in e.g.~\cite{Polyakov:1987ez,Strassler:1992zr,Schubert:2001he}. After providing a perturbative definition of the theory, we then focus on a simple class of amplitudes with on-shell CSP photons for the remainder of this paper. Though we briefly describe the more obvious factorization properties of amplitudes where an intermediate CSP photon goes on-shell, more detailed calculations and analysis will be left for future work. 

The results given below are encouraging for the consistency of the interacting theory defined by our rules, but do not constitute a proof even for general tree-level amplitudes, let alone at higher orders. Many questions will remain about which CSP vertex operators are consistent, potential anomalies, properties of loop amplitudes, and renormalizability, not to mention potential generalizations to non-abelian and gravitational theories. We hope our work represents a sharp step towards settling these questions in the interacting quantum theory of QED for $\rho\neq 0$ as a relatively simple test case. 

The work of \cite{Schuster:2023xqa, Schuster:2014hca} that we build on is one of several formalisms for describing states with $\rho\neq 0$.  Following several early attempts \cite{Yngvason:1970fy,Chakrabarti:1971rz,Iverson:1971hq,Abbott:1976bb,Hirata:1977ss,Zoller:1991hs}, multiple consistent worldline and field-theory formalisms for CSPs have been developed, including generalizations to higher dimensions, (anti-)deSitter backgrounds, fermions, and supersymmetric theories \cite{Bargmann:1948ck,Wigner:1963,Brink:2002zx,Edgren:2005gq,Khan:2004nj,Mourad:2005rt,Bekaert:2005in,Edgren:2006un,Boels:2009bv,Schuster:2013pta,Font:2013hia,Rivelles:2014fsa,Schuster:2014hca,Bekaert:2015qkt,Metsaev:2016lhs,Metsaev:2017ytk,Rivelles:2016rwo,Zinoviev:2017rnj,Alkalaev:2017hvj,Alkalaev:2018bqe,Buchbinder:2018soq,Buchbinder:2018yoo,Khabarov:2017lth,Metsaev:2017myp,Metsaev:2018lth,Najafizadeh:2017tin,Rivelles:2018tpt,Buchbinder:2019esz,Buchbinder:2019kuh,Metsaev:2019opn,Buchbinder:2019sie,Buchbinder:2020nxn,Burdik:2019tzg,Najafizadeh:2019mun,Metsaev:2021zdg,Buchbinder:2022msd,Najafizadeh:2021dsm}. Some of these have also been used to construct interactions \cite{Bekaert:2017khg,Bekaert:2017xin,Iverson:1971vy,Metsaev:2017cuz,Metsaev:2018moa}, though some of these are consistent only in the sense of BBVD currents \cite{Berends:1985xx}, in the limit of non-accelerating matter.  The  approach of \cite{Schuster:2023xqa} is of particular interest physically because the interactions are consistent for accelerating matter --- a prerequisite for constructing amplitudes --- and because they recover familiar physics of massless scalars or gauge fields at small $\rho$, and hence can plausibly extend to consistent descriptions of observed phenomena, in particular QED, or perhaps other fundamental forces.

{\it Notations and Conventions: } We use a $(+---)$ metric signature.  For a null four-vector $k^\mu$, we will often introduce a basis of null complex frame vectors $(k^\mu, q^\mu, \epsilon_+^\mu, \epsilon_-^\mu)$, where $\epsilon_+^* = \epsilon_-$, the only nonzero inner products are $\epsilon_+ \cdot \epsilon_- = -2$ and $q \cdot k = 1$.  The handedness of the basis is fixed by demanding $\epsilon^{\mu\nu\rho\sigma} = k^{[\mu} q^\nu \epsilon_+^\rho \epsilon_-^{\sigma]} / 2 i$, where $\epsilon^{0123} = 1$. As a concrete example, if $k^\mu = (\omega, 0, 0, \omega)$, we can choose
\begin{equation} \label{eq:frame_choice}
q^\mu = (1/2 \omega, 0, 0, -1/2 \omega), \qquad \epsilon_\pm^\mu = (0, 1, \pm i, 0).
\end{equation}
Eigenmodes of the helicity operator in a given reference frame, ${\v J}\cdot \hat {\v k}$, are constructed by choosing $\epsilon_\pm$ for each $k$ that satisfy $\epsilon_\pm^0 =0$ in that frame, or equivalently $\v{q} \propto \v{k}$.  
For brevity, we introduce the notation $[{\bar d}^4\eta] \equiv [d^4\eta] \delta'(\eta^2+1)$, where  $[d^4\eta]$ is the suitably regulated measure (e.g.~by analytic continuation to Euclidean space) defined in \cite{Schuster:2014hca, Schuster:2023xqa}.

\section{Defining Scalar QED for $\rho\neq 0$ in Perturbation Theory}
\label{Sec:pathIntegral}

We focus on scalar QED at nonzero $\rho$. The matter consists of a pair of charged scalar particles with mass $m$ and charge $\pm e$. We will label the matter states $\ket{p}$ by their momentum $p$ and suppress charge labels. The (Abelian) photon is a CSP with spin scale $\rho$, and its polarization states are labeled as $\ket{k,h}$. We remind the reader that the helicity polarizations are not in general Lorentz invariant. An infinitesimal Lorentz boost by $\v{v}_\perp$ in a direction transverse to $k$ mixes these states into each other, $U(\v{v}_\perp) \, |k, h \rangle = \sum_{h'} c_{hh'} |k', h' \rangle$ where the ISO(2) mixing coefficients for $h \neq h'$ are Bessel functions whose power series involves positive powers of $\rho \, v_\perp/k^0$. In other words, the spin scale characterizes the mixing of helicity modes under Lorentz transformations. This is analogous to what happens for ordinary massive particle under boosts, and has been discussed in detail in the literature (see~\cite{Schuster:2013pxj,Schuster:2013vpr}).  

\subsection{Review: Ordinary QED in the Worldline Formalism}
\label{ssec:pathIntegralQED}
Our starting point is a first-quantized formalism for (scalar) matter particles interacting with fields. Pioneered by Feynman in the case of quantum electrodynamics \cite{Feynman:1950ir} , this approach is less familiar to modern audiences than the second-quantized formalism of quantum field theory, but equivalent.  A modern resurgence of interest in this approach drew on the parallel with perturbative string theory and applied this formalism to derive the Bern-Kossower rules \cite{Strassler:1992zr,Strassler:1993km} and generating simple integral expressions for computations of multi-photon and graviton amplitudes (see e.g.~\cite{Ahmadiniaz:2019ppj,Ahmadiniaz:2021ayd}). The formalism, as applied to gauge theories, is reviewed in \cite{Strassler:1992zr, Schubert:2001he}. 
Our treatment mostly adopts a mix of the conventions of \cite{Ahmadiniaz:2019ppj} and \cite{Strassler:1993km}, but we work in Minkowski rather than Euclidean signature. 
To introduce our notation and formalism in a more familiar context, we begin by briefly reviewing ordinary Abelian scalar QED in this formalism, then extend to the context of CSP external states (and associated fields). 

A matter-matter correlator between $x$ and $x'$ in an arbitrary QED background field $A$ can be written as
\begin{align}
    \langle \phi(x) \bar\phi(x') \rangle_{A} = \int_{{\cal P}[x,x']} D{\cal E} Dz e^{i\int\limits_{\cal P} L(z,A,{\cal E})/\hbar}.  \label{WL_2point_QED}
\end{align}
Here the integral is over all paths ${\cal P}$ specified by functions $z(\tau)$ and ${\cal E}(\tau)$ over some finite range of the parameter $\tau$ that interpolate between $x$ and $x'$. The Lagrangian $L(z,A,{\cal E}) = \frac{1}{2{\cal E}} \dot z^2 +\frac{{\cal E}}{2} m^2 + i e \dot z \cdot A(z)$ for a relativistic particle in the background field $A^{\mu}(x)$ is invariant under worldline reparametrizations with an accompanying transformation of the einbein ${\cal E}$. Division of the path integral by the volume of the worldline reparametrization gauge group and an overall normalization of the path integral (see Appendix \ref{sapp:discretization}) are implicit. The integral $T \equiv \int d\tau {\cal E}(\tau)/2$ is invariant under worldline reparametrizations.  To fix gauge, we take ${\cal E}=2$ and choose paths to begin at $\tau=0$; the path integral over ${\cal E}$ then reduces to an integral over the total worldline time $T$, yielding
\begin{align}
    \langle \phi(x) \bar\phi(x') \rangle_{A} = \int_0^{\infty} dT \int\limits_{\substack{z(0)=x\\z(T)=x'}} Dz e^{i\int_0^{T} d\tau L(z,A)/\hbar}, \label{eq:startingPI}
\end{align}
with $L(z,A)=\frac{1}{4} \dot z^2+m^2 + i e\dot z \cdot A(z)$.  Note that in this gauge, $\tau$ has mass dimension -2 and for massive particles is related to the proper time by a factor of $2m$.  

Extracting connected tree amplitudes from the above requires a few manipulations.  Each outgoing photon leg of momentum $k_i$ and helicity $h_i$ must contract into one factor of $A(z)$ from the action.  Using the mode expansion of $A$, we see that this pulls down a factor of $\int_0^{T} dt_i V_{out}^{k_i,h_i}(t_i)$ with 
\begin{align}
 V_{out}^{k_i,h_i}(t_i) &= {\varepsilon_{h_i}^*}^\mu {{\hat V}_\mu}^{k_i}(t_i) \label{Vout_QED}\\
 {{\hat V}_\mu}^{k_i}(t_i) &= - e {\dot{z}}_\mu(t_i) e^{ik\cdot z(t_i)}. \label{Vhat_QED}
 \end{align}
Similarly, an incoming leg of momentum $k_j$ and helicity $h_j$ pulls down a factor $\int_0^{T} dt_j V_{in}^{k_j,h_j}(t_j)$
\begin{align}
V_{in}^{k_j,h_j}(t_j) = {\varepsilon_{h_j}}^\mu {{\hat V}_\mu}^{-k_j}(t_j). \label{Vin_QED}
\end{align}
Of course, these respect a crossing symmetry $V_{in}^{k,h}(t) = V_{out}^{-k,-h}(t)$ provided we choose polarizations satisfying $\varepsilon_+ = {\varepsilon_-}^*$ with no relative phase.  Beyond the powers of $A(z)$ that contract with external states to yield vertex operators, any additional powers of $A$ do not contribute to the tree amplitude for a single worldline, so we take $A=0$ in the Lagrangian, i.e.~$L(z) = \frac{1}{4} \dot z^2+m^2$. 

Since \eqref{eq:startingPI} has the interpretation of a matter-matter correlator, the procedure to take the matter on-shell is familiar: we first transform the worldline endpoints $x$ and $x'$ into Fourier space and then LSZ reduce. 
It is convenient to represent the Fourier factors as vertex operators as is done in \cite{Strassler:1993km}.  To do this, we first note that for any path length $T'$, the \emph{massless} free path integral $\int\limits_{-T'<\tau<0}^{z(0) = x_i} Dz e^{-\int_{-T'}^{0} d\tau \dot z^2/4}$  (with lower limit unconstrained) is a representation of the identity (and analogously for $T<\tau<T+T'$), as elaborated in Appendix \ref{sapp:identityForExtending}.  Using these identities, we can extend the path integral to go from $-T'$ to $T+T'$, recasting the Fourier factors for in- and outgoing matter momenta $p$ and $p'$ as vertex operators $V_{in}(0)$ and $V_{out}(T)$ on this extended path integral with 
\begin{align}
    V_{in}(t) \equiv e^{-ip\cdot z(t)},  \qquad V_{out}(t) \equiv e^{+ip'\cdot z(t)}. \label{eq:matterVtx}
\end{align} 

\begin{figure}[h]
\begin{tikzpicture}
 \tikzstyle{every node}=[font=\fontsize{8}{8}\selectfont]
    \begin{feynman}
    
\vertex[label=below:$p_0$] (neg)  at (-0.2,1) ;
\node[dot,label=below:{{$V_{out}^{p_0}$}}] (ti) at (1, 1); 
\node[dot,label=below:{{$V_{\gamma,out}^{k_1,h_1}$}}] (t1) at (2.2, 1);  
\node[dot,label=below:{$V_{\gamma,out}^{k_2,h_2}$}] (t2) at (3.8, 1); 
\node[dot,label=below:{$V_{out}^{p_3}$}] (tf) at (5, 1); 
\vertex[label=below:$p_3$] (pos) at (6.2,1);
\vertex[label={$k_1,h_1$}] (k1) at (2.2, 2.5);
\vertex[label={$k_2,h_2$}] (k2) at (3.8, 2.5);
\vertex[label={\small ${\cal M} = \int Dx \prod \left( dt_i V_i \right)  e^{\frac{i}{\hbar} \int d\tau \frac{1}{4}\dot z^2}$}] (caption) at (3.0, -1);

    \diagram* {
      (ti) -- [plain] (neg), 
      (ti) -- [plain] (tf),
      (tf) -- [plain] (pos), 

      (t1) -- [photon] (k1), 
      (t2) -- [photon] (k2), 
};


    \end{feynman}
\end{tikzpicture}
\qquad
\includegraphics[width=0.4\linewidth]{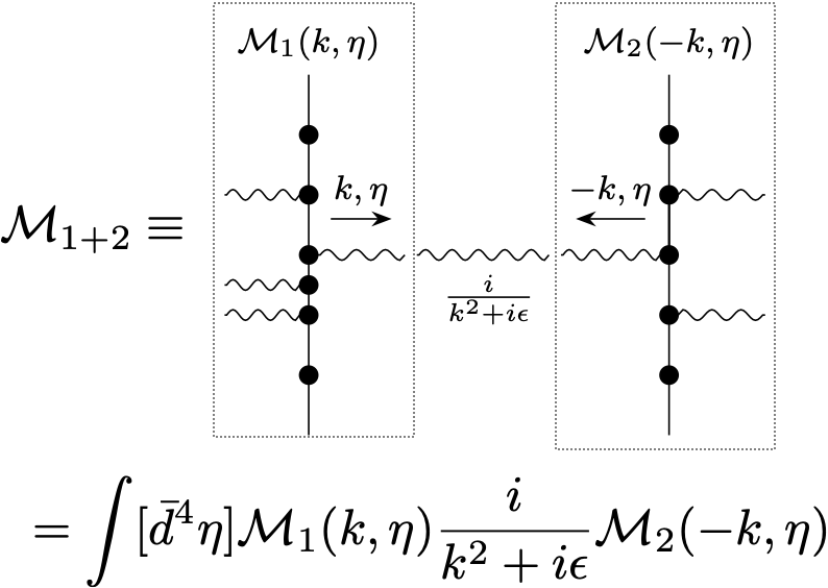} 
\caption{Left: On-shell CSP photon diagram topology with one matter leg and two insertions. Each black dot corresponds to a vertex operator on a free (massless) matter worldline. The vertex operators corresponding to external matter legs $p$ and $p'$ are required to be the first and last vertex, respectively; the intermediate insertions of CSP lines are unordered.  Below, a shorthand form of the corresponding integral expression (see Eq.~\eqref{eq:path_M_helTransitionElement}.  Right: Diagram and rules for an off-shell CSP photon diagram topology, with vertex operator $V^{k,\eta}$ given by \eqref{eq:offshell_vertex_operator}. Factorization when the CSP lime goes on-shell is discussed in Sec.~\ref{sec:unitarity}.
}
\label{fig:picsA}
\end{figure}

Thus, the tree amplitude with one in- and outgoing matter line and $n$ external photons (which, for brevity, we have taken to be all outgoing) is given by a path integral involving $n+2$ vertex operators at $\tau=0$, $\tau=T$, and $n$ arbitrary, unordered points in between:
\begin{align}
A(p,p',\{k_i,h_i \}) & = M(p,p',\{k_i,h_i \}) \bigg|_{LSZ} = \lim_{m^2} (p^2-m^2) ({p'}^2-m^2) M(p,p',\{k_i,h_i \}),
\label{eq:path_A_QED}\\
M(p,p',\{k_i,h_i \}) & =  \int_0^{\infty} dT e^{i m^2 T/\hbar} \int D z V_{in}^p(0) V_{out}^{p'}(T) \left(\prod_i \int dt_i V_{\gamma,out}^{k_i,h_i}(t_i)\right) \ e^{\frac{i}{\hbar} \int d\tau \frac{1}{4}\dot z^2}\label{eq:path_M_QED}
\end{align}
and the integral is now over extended paths of length $T+2T'$ with both endpoints of the path unconstrained (note that the above includes a momentum-conserving delta function, which arises from integration over arbitrary translations of each path).  It is convenient to further take $T'\to\infty$, translate $\tau$ to $\tau-t_1$, and reorder integrations with shifted labels to obtain the equivalent form 
\begin{align}
 M(p,p',\{k_i,h_i \}) & =  \int D z \int_0^{\infty} dT_i dT_f e^{i m^2 (T_i+T_f)/\hbar} V_{in}^p(-T_i) V_{out}^{p'}(T_f) \nonumber \\
 & \qquad\qquad \left(\prod_i \int_{-T_i}^{T_f} dt_i V_{\gamma,out}^{k_i,h_i}(t_i)\right) \delta(t_1) \ e^{\frac{i}{\hbar} \int d\tau \frac{1}{4}\dot z^2}\label{MfunctionQED}
\end{align}
This single integral sums over all Feynman diagrams in which external photons connect to the matter line; Figure \ref{fig:picsA} illustrates this schematically for a Compton amplitude, where the $s$ and $u$ channels correspond to distinct orderings $t_1<t_2$ vs $t_1>t_2$.  Notably, there is no two-photon vertex, only the Lagrangian coupling to the particle, $ie\dot z\cdot A(z)$; in the worldline computation, the $\varepsilon_1\cdot \varepsilon_2$  contribution to the amplitude will arise when the operators $V_{k1,h1}$ and $V_{k2,h2}$ are coincident.  

When the external polarizations $\varepsilon$ are chosen to be circularly polarized (i.e. proportional to one of the reference polarizations $\epsilon_\pm(k)$ defined in the conventions, up to a shift by $k^\mu$), the resulting amplitude should transform covariantly under the little group.  This covariance follows from the behavior of the vertex operators $V^{k,h}$. As discussed in Ref.~\cite{Schuster:2014hca}, one can show that $\epsilon_{\pm}^{\mu}(k)$ are eigenvectors of the Little Group rotation generator $R$ with eigenvalue $\pm 1$, whereas the little group translation operators $T_{\pm}$ shift the $\epsilon_{\pm}^{\mu}(k)$ into $k^{\mu}$; $k^\mu$ is annihilated by all little group generators by definition. 
Looking at the form of the vertex operators in \eqref{Vhat_QED}, the shift into $k^{\mu}$ yields an operator that is a total $t$-derivative. In summary, we have  
\begin{align}
R \ V^{k,h}(t) &= h \ V^{k,h}(t) \\
T_{\pm} \ V^{k,h}(t) &= 0 + \mbox{total t-derivative}.
\end{align}
The total time derivative term does not contribute to the LSZ reduced path integral, and so the amplitudes constructed with insertions of $V^{k,h}$ in the path integral satisfy the appropriate transformation rule for a massless state with momentum $k$, helicity $h$, and spin Casimir $\rho=0$. 

Finally, we introduce as a useful intermediate step the ``polarization-stripped'' M-function,
\begin{align}
M_{\mu_1\dots \mu_n}(p,p',\{k_i\})  = \int D z  \int_0^{\infty} dT_i dT_f & e^{i m^2 (T_i+T_f)/\hbar} V_{in}^p(-T_i) V_{out}^{p'}(T_f) \times \nonumber \\ & \quad\left(\prod_i \int_{-T_i}^{T_f} dt_i \hat V_{\mu_i}^{k_i}(t_i)\right) \delta(t_1) \ e^{\frac{i}{\hbar} \int d\tau \frac{1}{4}\dot z^2}, \label{polStripped_QED}
\end{align}
from which \eqref{MfunctionQED} is obtained by contracting in polarization tensors.   (Again, somewhat unconventionally, this ``M-function'' contains the overall momentum-conserving $\delta$-function).

We have already noted the crossing properties of the external photon vertices.  It is similarly straightforward to generalize this procedure to annihilation (pair-production) of a single particle-antiparticle pair into $n$ photons, simply by taking a negative-energy $p'$ ($p$), and to consider multiple external matter lines, each parametrized by its own worldline. 

Although we will not use it here, this formalism readily generalizes to include multiple worldlines at tree level, and an expansion in both matter and photon loops.  When there are multiple external worldlines present, each has its own path integral.  Likewise, higher orders in perturbation theory receive contributions from sectors with additional matter loops, treated as closed paths (see e.g.~\cite{Polyakov:1987ez,Strassler:1992zr}). In both loop diagrams and trees involving multiple matter legs, there are also contributions to the path integral where powers of $A$ from the action are contracted into one another, i.e. internal photon lines.  These can similarly be treated as a product of two $\hat V_\mu$ vertex operators (on the same or different worldlines), each with a free vector index, contracted by the usual gauge-field propagator, as discussed for example in \cite{Ahmadiniaz:2016qqo}. 

\subsection{Generalization to Non-Zero $\rho$}
Now that we have established notation in ordinary QED, it is straightforward to generalize to matter interacting with a CSP field.  The results of this subsection follow straightforwardly from including in the path integral Lagrangian the matter-CSP interaction that we considered in \cite{Schuster:2023xqa} in a classical context.  However, readers who are primarily interested in computing amplitudes can simply take the expressions in this section (in particular, vertex operator and tree amplitude formulas \eqref{eq:CSP_vertex_op_out}-\eqref{eq:path_M_helTransitionElement} and the CSP propagator rule \eqref{eq:CSP_vertex_op_out}) as definitions. 

In the local gauge field theory of \cite{Schuster:2014hca}, continuous spin particles are the quantized modes of a field $\Psi(x^\mu,\eta^\mu)$, where $\eta^\mu$ is an auxiliary four-vector.  It was shown in  \cite{Schuster:2023xqa} that the most general (Abelian) gauge invariant coupling of such a field to matter worldlines can be written as 
\begin{align}
L_{int}(\tau) & = \int d^4 x d^4k [{\bar d}^4\eta] \Psi(\eta,x) e^{-i k\cdot(x-z(\tau))} \sum_w g_w J_w(\eta,k,\dot z(\tau)) \label{L_CSP}\\
J_w(\eta,k,\dot z) &=  e^{-i\rho\frac{\eta\cdot\dot z}{k\cdot \dot z}} \left(\sqrt{2} i\, k\cdot \dot z/\rho\right)^w  + {\cal D} X(\eta,k,\dot z),\label{eq:Current}
\end{align}
where ${\cal D} =\left[ k^2  - (-ik\cdot \eta - \frac{1}{2}(\eta^2+1) (-ik\cdot \partial_\eta + \rho))\right]$ is the differential operator in the  free CSP equation-of-motion for the Fourier modes, $\Psi(k,\eta)$, of the CSP gauge field. The function $X$ is completely unconstrained by our present understanding. We will discuss some aspects of its physical interpretation later, but for now we will simply ignore it because the choice of $X$ does not at all impact amplitudes that involve only on-shell CSP legs.  To describe photons at non-zero $\rho$ we take $w=1$ and $g_1 = e$. For this case, as we show below, the helicity-1 mode of the CSP dominates interactions. More precisely, at energy scales large compared to $\rho$, this mode couples like a photon and all other helicity modes become decoupled. However, in the interest of completeness we will present rules here for arbitrary $w$. 

The rules for constructing CSP-matter amplitudes follow directly from replacing $ie \dot z \cdot A(z)$ with \eqref{L_CSP} in the worldline Lagrangian in \eqref{eq:startingPI}, and using the contraction rule (A32) of \cite{Schuster:2023xqa} to contract with external helicity wavefunctions.  We therefore have, in analogy with 
\eqref{Vhat_QED}, a vertex operator valid on and off-shell:  
\begin{align}
V_{\gamma, CSP}^{k,\eta}(t) \equiv i e\, e^{ik\cdot z(t)} \ \left( e^{-i\rho\frac{\eta\cdot \dot{z(t)}}{k\cdot \dot{z}(t)}} \ \left(\sqrt{2} i k\cdot\dot{z}(t)/\rho \right)^{(w=1)} +{\cal D} X(\eta,\dot{z(t)}) \right).
\label{eq:offshell_vertex_operator}
\end{align}
For \emph{on-shell CSP legs only}, the contributions from $X$ vanish so that we can simply use 
\begin{align}
\hat V_{CSP}^{k,\eta}(t) &\equiv i e\, e^{ik\cdot z(t)} \ e^{-i\rho\frac{\eta\cdot \dot{z(t)}}{k\cdot \dot{z}(t)}} \ \left(\sqrt{2} i k\cdot\dot{z}(t)/\rho \right)^{(w=1)},
\label{eq:CSP_vertex_operator_quotient_form}
\\
V_{CSP}^{out,k,h} &\equiv \int [{\bar d}^4\eta] \psi_h^*(\eta,k) \hat V_{CSP}^{k,\eta} = \int_0^{2\pi} \frac{d\phi_i}{2\pi} e^{-ih_i\phi_i} V_{CSP}^{k,\eta(\phi)} \label{eq:CSP_vertex_op_out}\\
V_{CSP}^{in,k,h} &\equiv \int [{\bar d}^4\eta] \psi_h(\eta,k) \hat V_{CSP}^{-k,\eta} = \int_0^{2\pi} \frac{d\phi_i}{2\pi} e^{ih_i\phi_i} V_{CSP}^{-k,\eta(\phi)},\label{eq:CSP_vertex_op_in}
\end{align}
where $\psi_h$ are wavefunctions defined in \cite{Schuster:2023xqa}, and the $\eta$-integral expressions can be simplified using identities in \cite{Schuster:2023xqa} to the form on the right-hand side with $\eta(\phi) \equiv \rm{Im}(e^{-i\phi} \epsilon_+(k))$ and $\epsilon_+(k)$ a reference polarization vector as defined in the conventions.\footnote{Throughout this work we use $\varepsilon$ to denote photon circular polarization vectors which satisfy $\varepsilon_+\cdot\varepsilon_- = -1$ and $\epsilon$ to denote frame vectors satisfying the conventions of \cite{Schuster:2023xqa} where $\epsilon_+\cdot\epsilon_- = -2$.} Each helicity vertex operator has the interpretation of a Fourier mode of $\hat V$, when $\eta$ is restricted to lie on a unit circle parametrized by $\phi$.  
Again these respect a crossing symmetry: $V_{CSP}^{in,k,h} = V_{CSP}^{out,-k,-h}$ (provided we define $\epsilon_+(-k) = \epsilon_+(k)$) so that we can freely choose an all-outgoing convention for the CSP legs. 

As before, the little group covariance of the scattering amplitudes, now for $\rho\neq 0$,  follows from the behavior of the vertex operators $V_{CSP}^{k,h}$. The vertex operators \eqref{eq:CSP_vertex_op_in} carry the same little group transformation properties as the wave-functions $\psi_h$. As shown in Ref.~\cite{Schuster:2014hca}, the little group generators for a null vector $k$ act on the wave-functions $\psi_h$ as $R \, \psi_h(\eta,k) = h \, \psi_h(\eta,k)$, and $T_\pm \, \psi_h(\eta,k) = \rho \ \psi_{h\pm 1}(\eta,k)$ up to pure gauge terms of the form ${\cal D}F(\eta,k)$. However, a pure gauge term of the form ${\cal D}F(\eta,k)$ replacing $\psi_h(\eta,k)$ in \eqref{eq:CSP_vertex_op_in} has vanishing $\eta$-integral. This crucially relies on the fact that the operators $\hat V_{CSP}^{k,\eta}$ in \eqref{eq:CSP_vertex_op_in} satisfy a Ward relation $\left( -ik\cdot\partial_{\eta}+\rho\right)\hat V_{CSP}^{k,\eta}=0$, generalizing the familiar QED Ward relation to the case $\rho\neq 0$. Thus, we have 
\begin{align}
R \ V_{CSP}^{k,h}(t) &= h \ V_{CSP}^{k,h}(t) \\
T_{\pm} \ V_{CSP}^{k,h}(t) &= \rho \ V_{CSP}^{k,h\pm 1}(t),
\end{align}
and so the amplitudes constructed with insertions of $V_{CSP}^{k,h}$ in the path integral satisfy the appropriate transformation rule for a massless state with momentum $k$, helicity $h$, and spin Casimir $\rho\neq 0$. 

In terms of these and the matter vertex operators, the ``polarization-stripped'' transition element analogous to \eqref{polStripped_QED} is 
\begin{align}
M(p,p',\{k_i,\eta_i\}) = \int D z \int_0^{\infty} dT_i dT_f & e^{i(T_i+T_f) m^2/\hbar} V_{in}^p(-T_i) V_{out}^{p'}(T_f) \times \nonumber\\
& \quad \left(\prod_i \int_{-T_i}^{T_f} dt_i \hat V^{k_i,\eta_i}(t_i)\right) \delta(t_1) \ e^{\frac{i}{\hbar} \int d\tau \frac{1}{4}\dot z^2}, \label{eq:path_M_polStripped}
\end{align}
which for $n$ external photons is now a function of $n$ free four vectors $\eta_i$, rather than a rank-$n$ tensor. What the $\eta$-space machinery of \cite{Schuster:2014hca,Schuster:2023xqa} facilitates is a straightforward generalization of familiar field theory expressions for transition elements to the case where physical particle states are {\bf not} in a finite dimensional representation of the Poincare Group, as occurs when $\rho\neq 0$. Though unfamiliar at first, once mastered, the reader will likely find this to be a compact and useful mathematical tool for making contact with field theory. 

Transition elements can be written (assuming all-outgoing convention for CSP legs) as 
\begin{align}
M(p,p',\{k_i,h_i\}) = \int_0^{\infty} dT_i dT_f &e^{i(T_i+T_f) m^2/\hbar} \int D z V_{in}^p(-T_i) V_{out}^{p'}(T_f)  \times \nonumber\\
& \quad \left(\prod_i \int_{-T_i}^{T_f} dt_i V^{k_i,h_i}(t_i)\right) \delta(t_1) \ e^{\frac{i}{\hbar} \int d\tau \frac{1}{4}\dot z^2}, \label{eq:path_M_helTransitionElement}
\end{align}
or equivalently in terms of the polarization-stripped element as
\begin{align}
M(p,p',\{k_i,h_i\}) = \int \left(\prod_i  \frac{d\phi_i}{2\pi} e^{ih_i\phi_i} \right) \hat V^{k_i,\eta_i}(t_i) M(p,p',\{k_i,\eta_i(\phi_i)\}).\label{eq:path_M_fromPolStripped}
\end{align}
Finally, amplitudes $A(p,p',\{k_i,h_i\})$ are obtained from $M(p,p',\{k_i,h_i\})$ above by LSZ reduction of the matter legs. 

Diagrams with loops or multiple matter lines can also have internal CSP lines. These are included (focusing on the tree case for clarity) by inserting an operator $V_{CSP}^{k,\eta}(t)$ for each worldline to which the CSP line attaches, and contracting the two $\eta$ indices. Specifically, two sub-diagrams $M_1( q,\eta,\dots)$ and $M_2(-q,\bar{\eta},\dots)$ connected by a single internal CSP line contribute to the total amplitude as
\begin{align}
M(1+2) = \int [{\bar d}^4\eta] M_1(q,\eta)\frac{i}{q^2+i\epsilon} M_2(-q,\eta), \label{eq:CSP_prop_rule}
\end{align}
where the measure and interpretation of the $\eta$-space integral is discussed in \cite{Schuster:2014hca, Schuster:2023xqa}. At this point, it should be far from obvious to the reader that this rule yields amplitudes that factorize into $M$-functions with on-shell CSP photons when $q^2=0$, but we'll explain how this works in Section \ref{sec:unitarity}.

In the context of internal CSP lines, we will find different results depending on the choice of the function $X$ appearing in the CSP vertex operator \eqref{eq:offshell_vertex_operator}. The freedom in $X$ is highly analogous to specifying the ``shape'' (radial profile, e.g.~charge radius, and non-rigidity) of a spherically symmetric particle whose CM motion is characterized by $z$. Such freedom exists also in QED, allowing the description of non-minimal couplings in addition to the minimal coupling assumed above --- but radiation emission/absorption depends only on the particle's total charge.  An important difference is that for CSPs, it is not yet clear whether $X=0$ or some other specific choice of $X$ should be viewed as ``minimal coupling'', or even what spacetime support for $X$ is physically sensible.  In particular, the first term in $J_w$, transformed back to position space, is not localized in spacetime to $x=z(\tau)$, and different choices of $X$ can change this spacetime profile --- see Sec.~IV of \cite{Schuster:2023xqa} for discussion. We hope that properties of amplitudes with intermediate CSP legs will further constrain the functional form of $X$.
At any rate, for a given choice of off-shell CSP vertex operator (i.e. choice of the arbitrary function $X$ in \eqref{eq:offshell_vertex_operator}), the above rules completely define the quantum theory in a perturbative expansion. These rules are manifestly Lorentz covariant, but at the cost of obscuring unitarity. 

A special class of amplitudes --- those involving matter propagators and an arbitrary number of on-shell external CSP legs, but no off-shell CSP legs --- are independent of the choice of $X$.  One can therefore derive results for this subset of amplitudes without settling the ``shape'' ambiguity in CSP currents noted in \cite{Schuster:2023xqa}.  In this paper, we will construct the simplest example of this class of amplitudes, in which two CSP photon legs couple to a single worldline, namely Compton scattering, comment on its physics, and illustrate its unitarity. By crossing, this computation will also yield the amplitude for pair annihilation into a pair of CSP photons.  


\subsection{Recovering QED Vertex Operators at High Energies}\label{ssec:rhoZeroLimit}
Above, it was stated that the current \eqref{eq:Current} and hence the vertex operators \eqref{eq:CSP_vertex_operator_quotient_form}-\eqref{eq:CSP_vertex_op_in} with $w=1$ approach QED-like interactions at energies large compared to $\rho$.  This will be seen clearly in the physical amplitude we consider in the next section. We sketch here the argument for why it is true more generally.  We begin by expanding \eqref{eq:CSP_vertex_operator_quotient_form} as a series in $\rho$:
\begin{align}
V^{k,\eta}_{\gamma,CSP}(t)  \approx e^{i k\cdot z(t)} \left( - \sqrt{2}\, e\, \frac{1}{\rho} k\cdot \dot z(t) + \sqrt{2} i\,e\, \eta\cdot \dot z(t) + {\cal O}(\rho)\right).
\end{align}
The second term is precisely $(- i \sqrt{2} \eta^{\mu}) {{\hat V}_{QED,\mu}^{k}}(t)$ where $\hat V_{QED}$ is the ordinary QED vertex operator \eqref{Vhat_QED}.  As discussed in \cite{Schuster:2023xqa}, QED can be embedded in $\eta$-space with the identification $\psi(\eta) = \sqrt{2}\eta^\mu A_\mu$ at $\rho=0$, so that this term in the CSP vertex operator precisely pulls out the QED vertex operator, up to an overall phase convention. 

The first term appears to diverge at small $\rho$, but is proportional to 
\begin{align}
k\cdot \dot z(t) e^{ik\cdot z(t)} = -i \partial_t e^{i k\cdot z(t)}. 
\end{align}
Comparing to \eqref{Vout_QED}, we see that this differs from a QED vertex operator by a replacement $\epsilon^\mu \to k^\mu$.  The Ward identity guarantees that a full on-shell amplitude evaluated with such a vertex replacement must vanish.  Thus, the coefficient of the naively divergent $1/\rho$ term vanishes and, in the $\rho\to 0$ limit, we recover precisely a QED matrix element.  
More formally, we also note that this vertex is a total $t$ derivative.  Because vertex operators $V(t)$ are always integrated over all possible $t$, this total derivative reduces to a boundary term evaluated at the worldline's physical endpoints $-T_i$ or $T_f$. It will turn out that these boundary terms lack one of the propagator singularities needed to survive LSZ reduction, and so these $1/\rho$ effects never contribute to physical amplitudes (though they will appear off-shell). 

An alternative, and equivalent, route to constructing amplitudes would have been to define 
\begin{align}
\hat V_{CSP}^{k,\eta}(t) & = i e^{ik\cdot z(t)} \ \left(e^{-i\rho\frac{\eta\cdot \dot{z(t)}}{k\cdot \dot{z}(t)}} -1 \right) \ \left(\sqrt{2} i k\cdot\dot{z}(t)/\rho \right).  
\end{align}
This form precisely recovers the scalar QED vertex operator as $\rho\to 0$, as the divergent term has been explicitly subtracted off (this expression has a convergent Taylor series in positive powers of $\rho$). Much like the QED vertex operator, it is not exactly conserved for off-shell matter but physical matrix elements evaluated between on-shell states \emph{are} always conserved.  The equivalence of this expression to the one we work with follows, again, from the QED Ward identity, which physically is just encoding the conservation of charge for the matter.  

\subsection{CSP Vertex Operators in Exponential Form}
The basic insight of the ``string-inspired'' approach is that path integrals of the form appearing in \eqref{eq:path_M_polStripped} can be solved exactly provided the vertex operators are sufficiently simple.  In particular, because the action is quadratic in $z$, the path integral is Gaussian provided that the vertex operators can be written in terms of exponentials in phases linear (or quadratic) in $z$ and $\dot z$. The matter vertex operators already have this form. 
For QED vertex operators, this is achieved by rewriting e.g. \eqref{Vout_QED} as
\begin{align}
    V^{k_i,h_i}_{out}(t_i) = i\,e \ \left(e^{{ik_i\cdot z(t_i) + \kappa_i \varepsilon_{h_i}^* \cdot \dot z(t_i)}}\right)_{\rm{lin.}\,\kappa_i},
\end{align}
where $\rm{lin.}\,\kappa_i$ denotes taking the piece linear in $\kappa_i$ at the end of the calculation.  
 
In a similar spirit, writing the CSP photon vertex operators in Gaussian form motivates the introduction of auxiliary variables to write $\hat{V}_{\gamma, CSP}^{k,\eta}(t)$ as 
\begin{align}
&\hat{V}_{\gamma,CSP}^{k,\eta}(t) = \int [d \mu] e^{i\newvar^\mu(\eta,k) \dot z_{\mu}(t)} \ e^{ik\cdot z(t)}
\label{eq:CSPvertexOperator}\\
&[d\mu] \equiv (\sqrt{2} i)^w \frac{d\theta}{\theta^w} d\lambda ds \ e^{-i \lambda \rho}\frac{d}{ds}\left(\frac{2 s}{s^2+\epsilon^2}\right)
\qquad \newvar^\mu \equiv \theta \eta^{\mu} - (s-\lambda\theta) k^{\mu}.
\label{eq:CSPvertexOperatorAuxiliaryDefinitions}
\end{align}
The \emph{existence} of some representation of the form \eqref{eq:CSPvertexOperator} suffices for the arguments that follow; the detailed definitions in \eqref{eq:CSPvertexOperatorAuxiliaryDefinitions} are inessential (and probably not unique).  Nonetheless, for a real four-vector $V$, it is readily verified that 
\begin{align}
\int [d\mu]  e^{i\newvar^\mu(\eta,k) \cdot V} &= (\sqrt{2} i)^w \int \frac{d\theta}{\theta^w} d\lambda ds \ e^{-i \lambda \rho}\frac{d}{ds}\left(\frac{2 s}{s^2+\epsilon^2}\right) e^{i( \theta \eta^{\mu} - (s-\lambda\theta) k^{\mu}) \cdot V} \\
& = (\sqrt{2} i)^w \int \frac{d\theta}{\theta^w} ds \ \frac{d}{ds}\left(\frac{2 s}{s^2+\epsilon^2}\right) e^{i( \theta \eta\cdot V - s k\cdot V) }\delta(\rho - \theta k\cdot V) \\
& =(\sqrt{2} i k\cdot V/\rho)^w \left( \frac{1}{|k\cdot V|} \int ds \frac{d}{ds}\left(\frac{2 s}{s^2+\epsilon^2}\right) e^{-is k\cdot V} \right)   e^{-i\rho \frac{\eta\cdot V}{k\cdot V}} \\ 
& = (\sqrt{2} i k\cdot V/\rho)^w e^{-i\rho \frac{\eta\cdot V}{k\cdot V}}, \label{eq:integralResultGeneralV}
\end{align}
so that \eqref{eq:CSPvertexOperator} reproduces \eqref{eq:CSP_vertex_operator_quotient_form}.  
Similar integrals involving imaginary $V$ will appear in the path integral, and can be \emph{defined} by analytically continuing the right-hand side \eqref{eq:integralResultGeneralV}.
\section{Examples: Compton Scattering and Pair Annihilation}
\label{sec:vecCompton}
The last section introduced a toolkit for computing CSP amplitudes in the worldline path integral formalism.  We now illustrate how this toolkit works in practice by setting up and solving the path integral for Compton scattering (or pair annihilation to two CSP photons, which is related by crossing).  As this is a physically interesting example, we will also study the behavior of the amplitude in several limits of physical interest, including the $\rho\to 0$ limit (Sec.~\ref{ssec:rhoZeroCompton}), the high-energy scaling, when $\rho$ is nonzero but parametrically smaller than kinematic invariants (Sec.~\ref{ssec:highEnergyCompton}), and the forward scattering limit $t\to 0$ (Sec.~\ref{ssec:forwardCompton}).  We defer discussion of other limits of physical interest --- in particular the low-energy limit for the case when the matter is massless --- to future work. 

\subsection{Solving the Path Integral}\label{ssec:solvingPI}
To illustrate the utility of this formalism, and in particular the exponential form \eqref{eq:CSPvertexOperator} of the vertex operator, we  focus for concreteness on a Compton-like amplitude $A(p_0;p_3;k_1,h_1; k_2,h_2)$ where the charged scalar worldline couples to two CSP vertices. We consider massless matter ($m=0$) for simplicity, though the calculation proceeds similarly with nonzero $m$.  We will work in all-outgoing conventions so that four-momentum conservation implies $p_0 + p_3 + k_1 + k_2 = 0$. The corresponding transition element $M(p_0;p_3;k_1,\eta_1; k_2,\eta_2)$ from which to derive this involves a product of the 4 vertex operators, 
\begin{align}
V_{out}^{p_0}(-T_i) V_{out}^{p_3}(T_f) \hat{V}_{\gamma,CSP}^{k_1,\eta_1}(0) \hat{V}_{\gamma,CSP}^{k_2,\eta_2}(t_2).
\end{align}
We can rewrite this product as a current-coupling
\begin{align}
 \int [d\mu_1] [d\mu_2] e^{- \int d\tau z(\tau) j(\tau)},
\end{align}
by defining
\begin{align}
j(\tau) & = \bar j(\tau) + \hat j(\tau)  
\label{eq:jdefinition}\\
\bar j_(\tau) &= i \left( p_0 \delta(\tau + T_i) + p_3 \delta(\tau - T_f) + k_1 \delta(\tau) + k_2 \delta(\tau-t_2)\right)\\
\hat j(\tau) &= i \left( \newvar_1 \delta'(\tau) + \newvar_2 \delta'(\tau-t_2)\right).
\end{align}
We note for future reference that $P=-i \int d\tau j(\tau) = p_0 + p_3+ k_1 + k_2$ is the sum of outgoing momenta, which is required to vanish by translation invariance. 

Following string inspired techniques, we will evaluate the path integral in Euclidean form, and then continue back to Minkowski signature near the final stage of the calculation. This is achieved by mapping worldline time coordinates to $\tau\rightarrow i\tau$, and similarly mapping $q^0\rightarrow iq^0$ for all momentum vectors $q$. The vectors $\eta_i$ undergo the same map, which is actually quite natural as Euclidean continuation is also used to regulate $\eta$-space integrations.  The path-integral form of the transition element is now given by 
\begin{align}
M={\int [d\mu_1] [d\mu_2] }   \int_0^\infty dT_i dT_f \int_{-T_i}^{T_f} dt_2 \int D z(\tau)
e^{-\int d\tau  \left(\frac{1}{4} \dot z(\tau)^2 + j(\tau)\cdot \dot z(\tau)\right)}
. \label{eq:2emissionPathIntegral}
\end{align}
This can be evaluated exactly by introducing a Green's function and completing the square using the translation-invariant Green's function 
\begin{equation}
G(\tau-\tau') \equiv \left(\frac12 \partial_\tau^2 \right)^{-1} = |\tau-\tau'|\label{eq:G_TI}
\end{equation}
(using any alternative Green's function yields the same results once boundary contributions are treated consistently).  The result is
\begin{align}
\int D z(\tau) e^{\frac12 \int d\tau  z(\tau) (\frac12 \partial_\tau^2) z(\tau) -2 j(\tau) z(\tau)} = (2\pi)^4 \delta^{(4)}(P) e^{-\frac12 \int d\tau d\tau' j(\tau)G(\tau-\tau') j(\tau')},\label{eq:PdeltaFunction}
\end{align}
where in the final expression the $(2\pi)^4 \delta^{(4)}(P)$ arises from integrating over overall space-time translations of the worldline. The origin of the $\delta$-function and the equivalence of results obtained using different Green's functions are both elaborated in Appendix \ref{sapp:BCZeroModeDelta}. We will suppress the $(2\pi)^4 \delta^{(4)}(P)$ factor in the following; the remaining ``amplitude'' depends only on the current contractions with the Green's function:
\begin{align}
M=\int [d\mu_1] [d\mu_2] \int\limits_{\substack{0<T_{i,f}<\infty\\ -T_i < t_2 < T_f}} 
dT_i dT_f dt_2 
e^{-\frac12 \int d\tau d\tau' j(\tau)G(\tau-\tau') j(\tau')}. \label{eq:2emissionPathIntegralSolved}
\end{align}

The evaluation of \eqref{eq:2emissionPathIntegralSolved} follows techniques that are standard in the literature on ``string-inspired'' worldline path integrals, but we evaluate them in some detail as they are likely unfamiliar to many readers. To simplify notation, we define \begin{align}
    j_1 \circ j_2 \equiv -\frac 12 \int d\tau d\tau' j_1(\tau)G(\tau-\tau') j_2(\tau').
\end{align}
We then have
\begin{align}
j\circ j & = \bar j \circ \bar j + 2 \bar j \circ \hat j + \hat j \circ \hat j \\
\bar j \circ \bar j & = \frac 12 \int d\tau d\tau' ( p_0 \delta(\tau + T_i) + p_3 \delta(\tau - T_f) + k_1 \delta(\tau) + k_2 \delta(\tau-t_2))|\tau-\tau'|\\
 &\qquad \qquad \times ( p_0 \delta(\tau' + T_i) + p_3 \delta(\tau' - T_f) + k_1 \delta(\tau') + k_2 \delta(\tau'-t_2)) \\
 &  = -p_0^2 T_i - p_3^2 T_f - t_2 (p_3 - p_0 - \sign\,t_2\; k_1)\cdot k_2, \label{eq:propagatorTerms}
\end{align}
where we have simplified e.g. $|T_f|=T_f$ and $|-T_i-t_2| = -T_i-t_2$ on the domain of integration, but the sign on the $k_1\cdot k_2$ contraction is undetermined as $t_2$ can have either sign.

Contractions involving $\hat j$ can be related by integration by parts to derivatives of $G(\tau-\tau')$, i.e. $-\partial_{\tau'} G = \sign(\tau-\tau')$ and $\partial_\tau \partial_{\tau'} G = - 2 \delta(\tau-\tau')$, yielding
\begin{align}
    2 \bar j \circ \hat j & = \int d\tau d\tau' ( p_0 \delta(\tau + T_i) + p_3 \delta(\tau - T_f) + k_1 \delta(\tau) + k_2 \delta(\tau-t_2)) \\
    & \qquad \qquad \sign(\tau-\tau') ( \newvar_1 \delta(\tau') + \newvar_2 \delta(\tau'-t_2)) \\ 
    & = \newvar_1 \cdot ( p_3 -p_0 + \sign\,t_2\;k_2) + \newvar_2\cdot( p_3 -p_0 -\sign\,t_2\; k_1). \label{eq:palpha_crossTerms}
\\
\hat j \circ \hat j & = \frac 12 \int d\tau d\tau' ( \newvar_1 \delta(\tau) + \newvar_2 \delta(\tau-t_2))) (-2 \delta(\tau-\tau')) ( \newvar_1 \delta(\tau') + \newvar_2 \delta(\tau'-t_2)) \\
& = -2 \delta(t_2) \newvar_1 \cdot \newvar_2,
\end{align}
where we have used the regularization prescription $\delta(\tau) \delta(\tau) \to 0$ to remove the (naively divergent) $\newvar_1^2$ and $\newvar_2^2$ terms, as discussed in Appendix \ref{app:distributions}.
Again following the logic of Appendix \ref{app:distributions}, the square and higher powers of $\delta(t_2)$ should be taken to vanish, so that 
\begin{align}
    e^{\hat j \circ \hat j} = e^{-2 \delta(t_2) \newvar_1 \cdot \newvar_2} = 1-2 \delta(t_2) \newvar_1 \cdot \newvar_2.\label{eq:jhatTimesjhat}
\end{align}

Combining these terms, we have
\begin{align}
M=\int [d\mu_1] [d\mu_2]  \int_{\substack{0<T_{i,f}<\infty\\ -T_i < t_2 < T_f}} &
dt_2 dT_i dT_f 
e^{-p_0^2 T_i - p_3^2 T_f - ( p_3 -p_0 - \sign\, t_2\; k_1)\cdot k_2\, t_2} \\
& e^{(\newvar_1+\newvar_2)\cdot (p_3 - p_0 ) + \sign(t_2) (\newvar_1\cdot k_2 -\newvar_2\cdot k_1)}
(1-2 \delta(t_2) \newvar_1 \cdot \newvar_2)\bigg|_{LSZ}
\end{align}
The dependence on $T_i$ and $T_f$ is simple, and can be further simplified by reversing the order of integration.  Integrating over $\max(0,\mp t_2) < T_{i,f} < \infty$ yields a factor of 
$\frac{1}{p_0^2} \frac{1}{p_3^2} e^{-t_2 (\theta(t_2) p_3^2 - \theta(-t_2) p_0^2)}$.  The first two factors are the external matter-leg propagators; the exponentials combine with the $t_2$-dependence of \eqref{eq:propagatorTerms} to yield the appropriate intermediate-state propagator depending on the sign of $t_2$.  The resulting integral is 
\begin{align}
M=\frac{1}{p_0^2\,p_3^2} \int [d\mu_1] [d\mu_2] e^{(\newvar_1+\newvar_2)\cdot (p_3 - p_0)} \int_{-\infty}^\infty 
dt_2 e^{- t_2 \left(s\,\theta(t_2) - u \,\theta(-t_2)\right)}  e^{ \sign(t_2) A_{12}}
\left(1-2 \delta(t_2) \newvar_1 \cdot \newvar_2\right), \label{eq:CSP_Compton_pre_t2Int}
\end{align}
where we have pulled out a common factor from the $t_2$ integral and introduced $A_{12} \equiv \newvar_1\cdot k_2 -\newvar_2\cdot k_1$, $s=(p_0+k_1)^2$, $u=(p_0+k_2)^2$.  LSZ reduction of the matter legs simply strips off the prefactor $\frac{1}{p_0^2\,p_3^2}$.

We now perform the $t_2$ integration, splitting the non-singular and singular terms. The non-singular part (i.e.~the piece proportional to 1 in the final parenthesis, not involving $\delta(t_2)$ is readily evaluated by separating the integral into domains with positive and negative $t_2$, and gives simply
\begin{align}
\frac{e^{A_{12}}}{s} + \frac{e^{-{A_{12}}}}{u}. \label{eq:nonsingPart}
\end{align}
The $\delta(t_2)$ part of \eqref{eq:CSP_Compton_pre_t2Int} must be evaluated with care, because the $\delta$-function is integrated against a function dependent on $\sign(t_2)$, which is discontinuous at the same $t_2=0$ point where the $\delta$-function is singular.  Fortunately, the Leibnitz rule implies a unique result for this integral as discussed further in Appendix \ref{app:distributions}:
\begin{align}
   -\newvar_1 \cdot \newvar_2 \frac{1}{A_{12}} \left(e^{A_{12}} - e^{-A_{12}}\right)
   =  -\newvar_1 \cdot \newvar_2  \int_{-1}^1 dx e^{x\,A_{12}}
   \label{eq:DeltaIntegral}
\end{align}
This result can also be obtained by regulating the discontinuity and divergence of $\sign$ and $\delta$, and is regulator-independent as long as $\frac{d}{dx} \sign(x) = 2 \delta(x)$ holds for the regulated forms --- which is clearly justified here, since our $\sign$ and $\delta$ arose from performing sequential derivatives on the same Green's function.  The second form of the result, as an integral over $x$, can be interpreted as ``stretching out'' the support of $\delta(t_2)$ into a finite region over which $\sign(t_2)$ varies linearly, but in which $|t_2|$ is sufficiently small that the expression $e^{-t_2 (\dots)}$ in \eqref{eq:CSP_Compton_pre_t2Int} is 1. 

 At this stage it is simple to analytically continue to Minkowski signature. Such a continuation requires that we specify a pole prescription.  The prescription we propose is to \emph{add $i\epsilon$ to the poles obtained from worldline-coordinate integrals of infinite extent}, i.e. the non-singular part \eqref{eq:nonsingPart}, \emph{but not in the terms involving $\newvar$}.  This prescription is compatible with unitarity requirements as discussed in Sec.~\ref{sec:unitarity}, produces the standard Feynman propagators for matter, and is logical when one motivates the introduction of $i\epsilon$ factors as ensuring convergence of \eqref{eq:CSP_Compton_pre_t2Int} when $t_2$ is continued to the imaginary axis, e.g.~for positive $t_2$,
 \begin{align}
\int_{0}^\infty 
dt_2 e^{- t_2 s} \rightarrow -i\int_{0}^\infty 
dt_2 e^{ it_2 (s+i\epsilon)}.
\end{align}
 It is less clear how one would define the integral of the singular term \eqref{eq:DeltaIntegral} when $t_2$ is taken to be imaginary, but it appears that no $i\epsilon$ term is needed, so we do not add one.  Other prescriptions may also be compatible with unitarity --- a question we defer to future work.  

Note that $e^{x\,A_{12}}$ from \eqref{eq:DeltaIntegral} and the $t_2$-independent prefactor $e^{(\newvar_1+\newvar_2)\cdot (p_3- p_0)}$ from \eqref{eq:CSP_Compton_pre_t2Int} group into the form
\begin{align}
    e^{\newvar_1\cdot P_1(x) + \newvar_2\cdot P_2(x)}
\end{align}
where 
\begin{align}
    P_1(x) = p_3 - p_0 + x\, k_2, \quad P_2(x) = p_3 - p_0 - x\, k_1. \label{eq:P12def}
\end{align}
Thus, the full result is
\begin{align}
M_{LSZ}=\int [d\mu_1] [d\mu_2] \left( \frac{1}{s+i\epsilon} e^{\newvar_1\cdot P_1^+ + \newvar_2\cdot P_2^+}
+ \frac{1}{u+i\epsilon}  e^{\newvar_1\cdot P_1^- + \newvar_2\cdot P_2^-} \right) 
 - \int_{-1}^{1} dx \newvar_1\cdot \newvar_2 e^{\newvar_1\cdot P_1(x) + \newvar_2\cdot P_2(x)}
 \label{eq:LSZ_unitarityform}
\end{align}
where we have introduced the shorthand $P_i^{\pm}= P_i(\pm 1)$. This form of $M_{LSZ}$ has the external matter propagators removed, but the on-shell conditions for the external momenta have not been used. We note that $P_i^\pm$ are precisely the momentum factors that would appear at the $k_i$ vertex in $s$- and $u$-channel Feynman diagrams, respectively. In this form, it will be straightforward to check unitarity, which we will return to shortly. But first, we simplify a bit more and study the behavior of the result. 

When $p_0$ and $p_3$ are  on-shell, $P_i^{\pm}$ satisfy
\begin{align}
P_2^+\cdot k_2 + p_3^2 & = - P_1^+\cdot k_1 + p_0^2 = s \\
P_2^-\cdot k_2 - p_0^2 & = - P_1^- \cdot k_1 - p_3^2 = - u,
\end{align}
so that the LSZ-reduced M-function can be rewritten as 
\begin{align}
M_{LSZ}= \int [d\mu_1] [d\mu_2] \int_{-1}^{1} dx \left( \frac{d}{dx} \frac{1}{k_2\cdot P_2(x)+i\epsilon x} - \frac{d}{d P_1(x)} \frac{d}{dP_2(x)} \right) e^{\newvar_1\cdot P_1(x) + \newvar_2\cdot P_2(x)},\label{eq:scalarAmplitudeSimplifiedBeforeMuIntegrals}
\end{align}
where we have rewritten the $\newvar_1\cdot\newvar_2$ prefactor as a differential operator to simplify the dependence on auxiliaries. 
This is the final result we will obtain in the auxiliary form.

The auxiliary integrals themselves reduce to the simple form 
\begin{align}
    \int [d\mu_i] e^{\newvar_i \cdot P_i(x)}.
\end{align}
This is the defining form of \eqref{eq:CSPvertexOperator}, up to a factor of $i$, and can be interpreted by analytic continuation from imaginary $P_i$, giving \footnote{This is entirely distinct from our earlier continuation between Euclidean to Minkowski signature}
\begin{align}
    \int [d\mu_i] e^{\newvar_i \cdot P_i(x)} \to \left(\sqrt{2} \,k_i\cdot P_i(x)/\rho\right)^{w} e^{-i\rho \frac{\eta_i\cdot P_i(x)}{k\cdot P_i(x)}}.\label{eq:naiveAlphaIntegrationWithoutI}
\end{align}
The isolated essential singularity in $k\cdot P_i$ remains on the real axis, which turns out be important for the unitarity of the final result. Taking $w=1$ for the vector-like coupling, and applying the above replacement to \eqref{eq:scalarAmplitudeSimplifiedBeforeMuIntegrals}, we obtain the M-function for scalar matter Compton scattering in integral form:
\begin{align}
M_{LSZ} = 2 \int_{-1}^{1} dx & \left( \frac{d}{dx} \frac{1}{k_2\cdot P_2(x)+i\epsilon x} - \frac{d}{d P_1(x)}\cdot \frac{d}{dP_2(x)} \right)
\nonumber \\ & \qquad \left(\frac{k_2\cdot P_2(x) k_1\cdot P_1(x)}{\rho^2} e^{-i\rho\frac{\eta_1 \cdot P_1(x)}{k_1\cdot P_1(x) } -i\rho\frac{\eta_2 \cdot P_2(x)}{k_2\cdot P_2(x) }}\right) \label{vectorMFunctionUnsimplified}.  
\end{align}
Each term above is separately of ${\cal O}(1/\rho^2)$, but a remarkable cancellation between the two terms yields a result of ${\cal O}(\rho^0)$. This is most easily exhibited after dropping $i\epsilon$'s, which we will do for the remainder of this discussion of tree amplitudes except for the discussion of unitarity in Sec.~\ref{sec:unitarity}.
One way to see this is by noting the operator relation (valid only when all external legs are on-shell)
\begin{align}
  \left [\partial_{P_1}\cdot\partial_{P_2}, \frac{k_2\cdot P_2 k_1\cdot P_1}{\rho^2}\right] 
& = \frac{d}{dx} \frac{k_1\cdot P_1(x)}{\rho^2},
\end{align}
which is actually related to the QED Ward identity.
We can then rewrite \eqref{vectorMFunctionUnsimplified} as
\begin{align}
M_{LSZ} &= -2 \int_{-1}^{1} dx \frac{k_2\cdot P_2(x)\, k_1\cdot P_1(x)}{\rho^2} \frac{d}{d P_1(x)}\cdot \frac{d}{dP_2(x)}  e^{-i\rho\frac{\eta_1 \cdot P_1(x)}{k_1\cdot P_1(x) } -i\rho\frac{\eta_2 \cdot P_2(x)}{k_2\cdot P_2(x)}} \label{vectorMFunctionsOperatorForm} \\
& = 2 \int_{-1}^{1} dx \left(\eta_1 - \frac{\eta_1\cdot P_1(x)}{k_1\cdot P_1(x)} k_1\right) \cdot \left(\eta_2 - \frac{\eta_2\cdot P_2(x)}{k_2\cdot P_2(x)} k_2 \right)  e^{-i\rho\frac{\eta_1 \cdot P_1(x)}{k_1\cdot P_1(x) } -i\rho\frac{\eta_2 \cdot P_2(x)}{k_2\cdot P_2(x).}}
\label{eq:VectorMFunctionSimplified}
\end{align}

This completes our computation of the $M$-function for Compton scattering generalized to nonzero spin Casimir $\rho\neq 0$, and illustrates how to perform path integral calculations with CSP photon vertex operators more generally. Next, we study the resulting amplitude, and the behavior for $\rho\rightarrow 0$ and more generally.  

\subsection{Standard Compton Amplitude in the Limit $\rho\to 0$}
\label{ssec:rhoZeroCompton}
The $\rho\to0$ limit of \eqref{eq:VectorMFunctionSimplified} is 
\begin{align}
2 \int_{-1}^{1} dx \left(\eta_1 - \frac{\eta_1\cdot P_1(x) k_1}{k_1\cdot P_1(x)}\right) \cdot \left(\eta_2 - \frac{\eta_2\cdot P_2(x) k_2}{k_2\cdot P_2(x)}\right).
\end{align}
This is homogeneous of degree 1 in each $\eta_i$.  Considering that each $\eta_i(\phi_i)$ in \eqref{eq:path_M_helTransitionElement} introduces one power of $e^{\pm i \phi_i}$, and helicity amplitudes are Fourier modes in $\phi_i$, this homogeneity implies that the amplitude is supported entirely in the helicity $\pm 1$ sector. Following \eqref{eq:CSP_vertex_op_out}, we find that helicity $h_j = \pm 1$ amplitudes are equivalent to replacing $\eta_j$ in the $M$-function by $\varepsilon_\pm/\sqrt{2}$, where $\varepsilon_\pm = (\mp i) \epsilon_\pm(k_j)/\sqrt{2}$ is a polarization vector satisfying the unit-norm condition $\varepsilon_\pm \cdot \varepsilon_\pm^* = -1$
\footnote{The first factor of $\sqrt{2}$ comes from the embedding of a  canonically normalized Maxwell field in the $\rho=0$ CSP field as the coefficient of $\sqrt{2}\eta^\mu$, as discussed in Sec.~2 of \cite{Schuster:2023xqa}. The second arises from the normalization convention $\epsilon_+\cdot \epsilon_- = -2$ (vs.~the unit norm condition satisfied by photon polarizations $\varepsilon_\pm$).  The factors of $\mp i$  simply switch to a more convenient choice of (arbitrary) phase for the helicity $\pm 1$ modes.}.
More generally, for any unit-norm photon polarization vectors $\varepsilon_j$, we can substitute $\eta_j\to \varepsilon_j/\sqrt{2}$. 
Thus, for any photon polarization vectors  $\varepsilon_1$ and $\varepsilon_2$,
\begin{align}
M^{V}_{\rho=0} = \int_{-1}^{1} dx \left(\varepsilon_1 - \frac{\varepsilon_1\cdot P_1(x) k_1}{k_1\cdot P_1(x)}\right) \cdot \left(\varepsilon_2 - \frac{\varepsilon_2\cdot P_2(x) k_2}{k_2\cdot P_2(x)}\right).\label{eq:integralFormScalarQED}
\end{align}
To see that this is indeed a rewriting of the Compton amplitude, we expand the product and use kinematic identities to group the $x$-dependent terms into a total $x$-derivative:
\begin{align}
M^{V}_{\rho=0} &= \int_{-1}^{1} dx\, \left( \varepsilon_1 \cdot \varepsilon_2 - \frac{d}{dx}\left[\frac{\varepsilon_1\cdot P_1^x\, \varepsilon_2\cdot P_2^x }{r(x)}\right]\right) \\
 &= 2 \varepsilon_1 \cdot \varepsilon_2 - \frac{\varepsilon_1\cdot P_1^+\, \varepsilon_2\cdot P_2^+ }{s} - \frac{\varepsilon_1\cdot P_1^-\, \varepsilon_2\cdot P_2^-}{u}
\end{align}
where $r(x) \equiv k_2\cdot P_2(x)$ interpolates between $s$ at $x=1$ and $-u$ at $x=-1$. These are precisely the contact, $s$-channel, and $u$-channel Feynman diagrams for Compton scattering in scalar QED, where our integral expression resembles a Feynman parametrization of the product of propagators\footnote{In pair-annihilation kinematics, a prescription must be specified for integrating \eqref{eq:integralFormScalarQED} across the singularity of the integrand at $r(x)=0$ --- a subtlety avoided for Compton scattering where $r>0$ throughout hte integration region. We have not explored this.}.

Our form of the amplitude has a few other amusing properties: the $x$-integrand is a simple product, where each factor $\left(\varepsilon_j - \frac{\varepsilon_j\cdot P_j(x)}{k_j\cdot P_j(x)}\right)$ is clearly invariant under gauge transformations $\varepsilon_j \to \varepsilon_j + \kappa k_j$.  Indeed, each factor is naturally rewritten as $\mp i F_j^{\mu\nu} {P_{j,\nu}(x)}/r(x)$ (with $-$ sign for $j=1$ and $+$ sign for $j=2$), where $F_j^{\mu\nu} = -i \varepsilon_j^{[\mu}k_j^{\nu]}$.  This makes gauge invariance even more manifest, but at the expense of locality, which is far from obvious. Another salient property of the Compton amplitude is also obscured in this form: the vanishing of the $++$ or $--$ helicity amplitudes.  Indeed, our $x$-integrand \emph{does not} vanish for this polarization choice.  That the amplitude does vanish can be reduced, in a convenient gauge where $\varepsilon_i\cdot k_j = 0$,  to a Feynman-parameter-like identity $\int_{-1}^1 dx \left( \frac{1}{su} +\frac{1}{r(x)^2}\right) = 0$. 

\subsection{Small-$\rho$ Expansion and Helicity Correspondence in Amplitudes}
\label{ssec:highEnergyCompton}
At $\rho=0$, the Compton amplitude is supported on ${h_1 = \pm 1, h_2 = \mp 1}$; all other modes decouple, and the theory reduces to scalar QED. At non-zero $\rho$, all helicities can interact --- but the amplitudes exhibit the same ``helicity correspondence''  observed in soft factors \cite{Schuster:2013vpr} and classical radiation and force-laws \cite{Schuster:2023xqa}: amplitudes involving ``partner'' polarization modes $h_i \neq \pm 1$  are suppressed by $n_1+n_2$ powers of $\rho$ over kinematic invariants, where $n_i = \left||h_i|-1\right|$.  The integral expression \eqref{eq:VectorMFunctionSimplified} already shows evidence of this structure: amplitudes for helicity $h_1 = n_1+1$ must involve at least $n+1$ powers of $\eta_1$, and all but one of these must arise from the phase, which generates an associated power of $\rho$. 

The most compact form of the amplitude is as an integral expression, which however somewhat obscures both the analytic structure of the result and its parametric scaling.  To address both of these points more concretely, we evaluate the integral \eqref{eq:VectorMFunctionSimplified}.  To do this simply, it is useful to introduce some new notation: $\tilde \eta_1 \equiv \eta_1 - \frac{\eta_1\cdot k_2}{k_1\cdot k_2} k_1$ and similarly for $\eta_2$,  so that $\tilde \eta_1 \cdot k_2 = \tilde \eta_2 \cdot k_1 = 0$, as well as $P=p_3 - p_0$ and $w \equiv \rho (\tilde\eta_1-\tilde\eta_2)\cdot P$.  Physically, $\tilde \eta$'s correspond to spatial polarization directions in the ``brick-wall'' frame of the two CSP photons (i.e., for our all-outgoing-CSP convention, $k_1 = ( - k_2^0, \v{k2})$).  We can write the general $M$-function as
\begin{align}
M_{LSZ} &= 2 e^{- i\rho \frac{\eta_1\cdot k_2 + \eta_2\cdot k_1}{k_1\cdot k_2}} \int_{-1}^{1} dx \left(\tilde\eta_1\cdot \tilde\eta_2 - \frac{\tilde\eta_1\cdot P \, \tilde\eta_2\cdot P t}{2 r(x)^2} \right)  e^{i\rho w/r(x)} \\ 
& = \left(- \frac{4\, 
\tilde\eta_1\cdot \tilde\eta_2}{t} \left[ r e^{i w/r} + i w E_1(-i w/r)\right]_{r=-u}^s + 2 i \tilde\eta_1\cdot P \tilde\eta_2\cdot P \frac{e^{i w/s} - e^{-i w/u}}{w}\right) e^{- i\rho \frac{\eta_1\cdot k_2 + \eta_2\cdot k_1}{k_1\cdot k_2}}.
\label{eq:VectorMFunctionBrickWall}
\end{align}
Before expanding in $\rho$, we consider briefly the analytic properties of the above expression. 
$E_1(z)= \int_z^{\infty} e^{-t}/t\, dt$ has a branch cut on the negative real axis, but is regular for $|\arg(z)|<\pi$, including on the imaginary axis where we evaluate it. The first term above naively appears to have unphysical singularities as $t\to 0$ (first term) or $w\to 0$ (second term), but in fact is regular in both limits.  As $t\to 0$, the upper and lower limits in the square-brackets expression converge, so that the first term evaluates to a derivative, approaching $ - 4\, \eta_1\cdot \eta_2 e^{i w/s}$ as $t\to 0$.  Meanwhile, as $w\to 0$, the ratio $\frac{e^{i w/s} - e^{-i w/u}}{w}$  approaches $i (1/s+1/u)$ as can be seen by Taylor-expanding the numerator. Therefore, despite appearances, the $t,w\to 0$ limits are smooth; the only singularities are the physical ones at $s,u\to 0$, i.e. those associated with an intermediate matter leg going on-shell for some ordering of CSP insertions.  

To better understand the $\rho$-scaling of Compton amplitudes at energies large compared to $\rho$, we can series-expand in $\rho$ to obtain, up to 2nd order in $\rho$,
\begin{align}
  M_{LSZ} &= M_0 + \rho M_1 + \rho^2 M_2 + \dots\label{eq:Mtaylor} \\ 
  M_0 &= 2 \left( 2 \eta_1\cdot \eta_2 - \eta_1\cdot P \eta_2\cdot P \left( \frac{1}{s} + \frac{1}{u}\right)\right) \\ 
  M_1 &= iw \left( 4 \eta_1\cdot \eta_2 \frac{\log\left(\frac{s}{-u}\right)}{t} + \frac{t}{s u} \left(\frac{1}{u}-\frac{1}{s}\right) \eta_1\cdot P \eta_2\cdot P \right)\\
  M_2 & = - w^2 \left( 2 \frac{\eta_1 \cdot \eta_2}{s\,u} + \frac{1}{3} \eta_1\cdot P \eta_2\cdot P \left(\frac{1}{u^3}+ \frac{1}{s^3} \right) \right).
\end{align}
This series expansion is only valid for $\rho$ small compared to the kinematic invariant $s,t,u$. The $\log$ term in $M_1$ arises from the $\log z$ term in the series expansion for $E_1(z)$; the coefficients of all higher orders in $\rho$ are rational functions.
Up to the factor of 2 explained earlier, $M_0$ has the structure of the scalar QED Compton amplitude. The higher-order poles in individual $M_i$ are misleading (approaching e.g. the point $u=0$ implies that $\rho$ is no longer small enough to use as an expansion parameter).  The behavior at small values of kinematic invariants therefore invovles resumming all terms of \eqref{eq:Mtaylor}, or more aptly working directly with \eqref{eq:VectorMFunctionBrickWall}).  We will see this most concretely in the forward-scattering limit below.

By counting $\eta$'s, and again noting that each factor of $\eta_i$ contributes $\pm \phi_i$ to the phase, we can see that $M_1$ will couple $h=\pm 1$ to $h=0,\pm 2$ modes at ${\cal O}(\rho)$, while 
$M_2$ contributes ${\cal O}(\rho^2)$ corrections to amplitudes in the helicity $\pm 1$ sector as well as providing the leading interaction for helicity pairs that differ by two units from the dominant pair of modes ${\pm 1,\mp 1}$, e.g.~$\{1,1\}$, $\{1,-3\}$, $\{0,2\}$, $\{0,0\}$, etc.  This is precisely the ``helicity correspondence'' scaling pattern anticipated above. A similar structure persists in higher-point amplitudes obtained from these rules, with the vanishing of negative powers of $\rho$ and matching of the standard QED amplitude at ${\cal O}(\rho^0)$  guaranteed by the QED Ward identity and series expansion of the vertex operator, as discussed in Sec.~\ref{ssec:rhoZeroLimit}. 

\subsection{Forward Scattering Limit}
\label{ssec:forwardCompton}
The bounded behavior of the amplitude for generic regions of phase space at $\rho\neq 0$ can be derived from the form given in \eqref{eq:VectorMFunctionBrickWall}. But a simpler limiting case to study is forward scattering where $p_0\rightarrow - p_3$, $k_1\rightarrow -k_2$.  In this kinematics, $t= 0$, $s= -u = 2k_1\cdot p_0$, and $\eta_1(\phi_1)\cdot k_{1,2}=0$ and  $\eta_2(\phi_2)\cdot k_{1,2}=0$. These relations simplify the integrand of \eqref{eq:VectorMFunctionSimplified} exactly, so that we find
\begin{align}
A(s;\phi_1,\phi_2) = 4\eta_1(\phi_1)\cdot\eta_2(\phi_2)e^{2i\rho \frac{(\eta_1(\phi_1)- \eta_2(\phi_2))\cdot p_3 }{s}} 
\label{eq:ForwardCompton}
\end{align}
In the angle basis, the amplitude is clearly bounded in both the soft and high energy limits, so that the forward total cross-section, which can be obtained by averaging $|A|^2$ from \eqref{eq:ForwardCompton} over $\phi_{1,2}$, is also clearly well-behaved in both the high- and low-energy limits. 

The Fourier transform \eqref{eq:path_M_fromPolStripped} yields helicity amplitudes that are sums of products of Bessel functions (up to a phase proportional to $(h_1-h_2)$ and related to the orientation of $p_3$ relative to the $\phi=0$ axis). Amplitudes with helicity $h_1 = h_2$ (where here $h_1$  is the physical \emph{incoming} helicity, and $h_2$ an outgoing helicity --- these correspond to $h_1 = - h_2$ in our all-outgoing conventions) are especially simple, evaluating to 
\begin{align}
A(s,h,h) = 2 \left(J_{h+1}^2(\rho \wp/s) + J_{h-1}^2(\rho \wp/s)\right),\label{eq:AmpForward}
\end{align}
where $\wp = |p_3|_\perp = |\v{p_3}\times \hat k_1|$ is the magnitude of the projection of $p_3$ onto the plane spanned by the $\eta(\phi)$.  At any nonzero $\rho$, the helicity amplitudes just approach the familiar scalar QED result in the high energy limit, and vanish in the deep infrared! 

Note that, for forward scattering in the CM-frame, $\wp =0$ identically so that \\ \mbox{$A(s,h,h) = 2 (\delta_{h,1} + \delta_{h,-1})$}. In other words, at any energy, in the CM frame forward scattering is supported only for $h = \pm 1$. Of course, since we are dealing with continuous spin states as external legs, their helicities transform under the little group and therefore under Lorentz transformations.  The more general result \eqref{eq:AmpForward} can be understood as the result of boosting the very simple CM-frame amplitude to an arbitrary frame, inducing a corresponding little-group transformation of the external legs.

 We briefly note that the amplitude is also remarkably simple in the  back-scattering $u\to 0$ limit, where it reduces to $4\tilde\eta_1\cdot \tilde \eta_2$ in the center-of-mass frame, just like \eqref{eq:ForwardCompton}.  
 It is easiest to see this by starting from \eqref{eq:VectorMFunctionBrickWall} directly. It is convenient to work in the brick-wall frame, where the overall phase in \eqref{eq:VectorMFunctionBrickWall} vanishes and $\tilde \eta_i(\phi_i) = \eta_i(\phi_i)$. This frame approaches the center-of-mass frame as $u\to 0$ (whereas as $t\to 0$ going from the CM to brick-wall frame requires an infinite boost).  Some care must be taken in approaching $u=0$, because the polarization-dependent $\tilde\eta_i\cdot P$ and $w$ also vanish as $\sqrt{u}$.  Thus, $iw/s$ vanishes as $u\to 0$ while $iw/u$ diverges, with a sign dependent on the spatial orientations of $\eta_{1,2}$.  
 However, using the boundedness of $e^{-ix}$ as $x\to \pm \infty$ and the limiting behaviors of $E_1$ ($\lim_{x\to \pm \infty} E_1(i x) =\pm i\pi$, $\lim_{x\to 0} x E_1(i x) = 0$), we recover the limiting form given above. 

\section{Remarks on Tree Level Unitarity and Factorization}
\label{sec:unitarity}
A general proof of unitarity at all orders in perturbation theory is beyond the scope of this paper. But at tree level, and for on-shell CSP photon amplitudes, unitarity works essentially the same way that it does in familiar scalar QED. We illustrate this with the Compton amplitude. At the level of $M$-functions, unitarity of amplitudes will be satisfied provided 
\begin{align}
& i \left( M(p;-p';\{k,h\};\{-k',-h'\}) - M(-p;p';\{-k,-h\},\{k',h'\})^* \right) \\
& \qquad = \sum_X  M(X;p',\{k',h'\})^*M(-p,\{-k,-h\};X)
\label{eq:unitarity}
\end{align}
We use this quantity rather than an on-shell amplitude because it is defined for off-shell $p$ and $p'$, where both sides of the equation are non-trivial.  In contrast, for Compton-scattering \emph{amplitudes} at real momenta with all legs on-shell, the $s$- and $u$-channel poles have vanishing coefficients and the right hand side also vanishes. 
The criterion of unitary $M$-functions for off-shell, real matter momenta is a non-trivial check, sufficient (but not strictly necessary) for the unitarity of amplitudes.  The logic we present here should generalize to higher-point M-functions and on-shell amplitudes.  We therefore sidestep the ambiguity in defining complex-momentum continuations of CSP amplitudes, which would be needed for the study of tree-level generalized unitarity along the lines of \cite{Benincasa:2007xk}. 

Consider the $s$-channel piece of the Compton $M$-function from \eqref{eq:LSZ_unitarityform}. Its contribution to the left of \eqref{eq:unitarity} is
\begin{align}
\int [d\mu_1] [d\mu_2] 2\pi\delta(s) e^{\newvar_1\cdot P_1^+ + \newvar_2\cdot P_2^+} 
\end{align}
To derive this, note that a change of variables $\theta\rightarrow -\theta, \lambda\rightarrow -\lambda$ in the auxiliary integration variables appearing in the expression for $M(p',\{k',h'\};-p,\{-k,-h\})^*$ keeps the $\rho$-dependent $e^{\newvar_i\cdot P_i}$ factors the same between both $M(p',\{k',h'\};-p,\{-k,-h\})^*$ and $M(p,\{k,h\};-p',\{-k',-h'\})$. This is precisely matched by the single particle contribution in the sum on the right hand side of \eqref{eq:unitarity}, where $X$ is a matter state with momentum $k_1+p$. The $u$-channel contributions work analogously, except now the $X$ contribution on the right hand side of \eqref{eq:unitarity} is a 3-particle intermediate state with 2 CSP photons and a matter state, where one of the CSP photon states is disconnected.  Finally, the contact term in \eqref{eq:LSZ_unitarityform} does not contribute to the LHS of \eqref{eq:unitarity}.  

The above logic should generalize to ``matter-leg'' cuts in more complex tree amplitudes.  What about intermediate CSP legs?  One prerequisite to unitarity is that, as a CSP of momentum $k$ goes on-shell, amplitudes should have a propagator pole $1/k^2$ whose residue is a sum over intermediate helicity states of factorized amplitudes.
Our rules \eqref{eq:CSP_prop_rule} for intermediate CSPs yield precisely such a pole, with residue
\begin{align}
\lim_{k^2\rightarrow 0} k^2M(1+2) &= \int [{\bar d}^4\eta] M_1(k,\eta) M_2(-k,\eta)  \label{eq:CSP_factorization_LHS}
\end{align}
where $M_i(k,\eta)$ are matter path integrals (or products thereof) with appropriate vertex operators inserted, including in particular an off-shell CSP vertex operator of the form \eqref{eq:CSP_vertex_operator_quotient_form}. 
Factorization requires that \eqref{eq:CSP_factorization_LHS} equal the product 
\begin{align}
\sum_h & M(1;\{k,h\})M(2:\{-k,-h\})\\
& =\sum_h\left( \int [{\bar d}^4\eta] M_1(\eta,k)\psi_h(\eta,k)^{\dagger}\right)\left(\int [{\bar d}^4\bar\eta]M_2(\bar{\eta},-k)\psi_h(\bar{\eta},k) \right) \\
& =
\int [\bar d\eta] [\bar d\bar{\eta}] M_1(\eta,k)M_2(\bar{\eta},-k)\sum_h\psi_h(\eta,k)^{\dagger}\psi_h(\bar{\eta},k),\label{eq:CSP_factorization_RHS}
\end{align}
where $M(i;\{k,h\})$ are \emph{almost} the same path integral expressions as $M_i(k,\eta)$, except that they involve on-shell CSP vertex operators \eqref{eq:CSP_vertex_op_in} and \eqref{eq:CSP_vertex_op_out} on the leg of momentum $k$.  

The final forms of the residue  \eqref{eq:CSP_factorization_LHS} and sum over products of sub-amplitudes \eqref{eq:CSP_factorization_RHS} involve the same product $M_1(\eta,k) M_2(\bar\eta,-k)$, except that the first expression has them evaluated at $\bar\eta=\eta$ while the latter contracts each into a wavefunction.  
Factorization therefore amounts to the remarkable relation 
\begin{equation}
    \int [\bar d\eta] M_1(\eta,k)M_2(\eta,-k) = \int [\bar d\eta] [\bar d\bar{\eta}] M_1(\eta,k)M_2(\bar\eta,-k) \sum_h\psi_h(\eta,k)^{\dagger}\psi_h(\bar{\eta},k).
\end{equation}
This relation was shown to hold in \cite{Schuster:2023xqa} provided each $M_i$ factor satisfies the (Ward identity) condition 
\begin{equation}
    \delta(\eta^2+1)\left(-ik.\partial_{\eta}+\rho\right)M_i(\eta,k) = 0.
\end{equation}
This condition follows trivially from our diagrammatic rules: the $\eta$-dependence of $M_i$ comes entirely from the vertex operator \eqref{eq:CSP_vertex_operator_quotient_form}, which itself manifestly satisfies this condition.  This makes clear that amplitudes involving intermediate CSP legs factorize correctly when the CSP leg is brought on-shell.  Besides factorization, tree-level unitarity also requires that the \emph{only} contributions to the left-hand side of the optical theorem are associated with intermediate propagators going on-shell.  We showed this above for our simple four-point M-function involving only matter propagators.  That it holds more generally, including in amplitudes with intermediate CSP propagators, is very plausible but not yet rigorously established. We leave this important question for future study. 

\section{Conclusion}
We have presented rules for computing scattering amplitudes in scalar QED where the photon is a CSP with non-zero spin invariant (2nd Poincare Casimir) $\rho\neq 0$. We illustrated how to use these general rules by calculating simple four-particle amplitudes. The results reduce to standard scalar QED Compton and pair annihilation amplitudes in the $\rho\to0$ limit, further generalizing the correspondence found in \cite{Schuster:2013vpr,Schuster:2023xqa}. Generalized Ward identities appropriate for nonzero $\rho$ ensure that amplitudes obtained from our rules factorize correctly, as required by unitarity. The approach given here should readily generalize to Yukawa-like theories, and it should be possible to generalize to General Relativity at leading order in Newton's constant. Generalizations to fully non-linear theories including Yang-Mills and GR represent an exciting direction of study, as does a more systematic study of anomalies, renormalizability, the UV and IR structure, and finding a more transparent and useful space-time interpretation of the CSP vertex operators. 

\subsection*{Acknowledgments}
We thank Lance Dixon and Kevin Zhou for general discussions and feedback, and Aidan Reilly for valuable feedback on the first arxiv version. The authors are supported by the U.S. Department of Energy under contract number DE-AC02-76SF00515. 
\appendix

\section{Expanded Discussion of Worldline Path Integrals}
\label{app:pathIntegral}
In Sections~\ref{ssec:pathIntegralQED} and \ref{ssec:solvingPI}, we stated several elementary path integral results without proof.  Though most of these are derived in one form or another in e.g.~\cite{Schubert:2001he,Strassler:1992zr} or references therein, we include their derivations here to provide a  self-contained reference with consistent notation throughout.
Section \ref{sapp:discretization} defines the path integral via discretization. This discretization (or some other regulator) is necessary to define and carefully account for the path integral's normalization.  Section \ref{sapp:identityForExtending} justifies the resolution of identity \eqref{eq:path_M_QED} used in Section~\ref{ssec:pathIntegralQED} to extend the massless path integral beyond the insertions of matter initial/final state wavefunctions. The extension procedure, inspired by the treatment of \cite{Strassler:1992zr}, allows for a more succinct and uniform treatment of all external legs as  matter and external-field legs. Finally, Section \ref{sapp:BCZeroModeDelta} derives \eqref{eq:PdeltaFunction}, including in particular the momentum-conserving $\delta$-function that arises from treatment of boundary terms (often associated in the literature with zero-modes of $\frac{1}{2} \partial_\tau^2$).  This section also discusses how the same result arises when using more general Green's functions than \eqref{eq:G_TI}. 

\subsection{Discretization and Free Path Integral Normalization}\label{sapp:discretization}
One of the simplest ways to carefully define and normalize the path integral between $x$ and $x'$ is by discretizing it into $N$ steps (i.e. introducing $N-1$ intermediate time points). Evaluating the resulting discretization of \eqref{eq:startingPI} in the free case, and demanding that it recover the standard two-point function, fixes the normalization.  

Concretely, for fixed path length $T$ and in $D$ space-time dimensions we define the discretized path integral as 
\begin{align}
    (Dz)_N = {\cal N}_N \prod_{j=1}^{N-1} d^D z(\tau_j),
\end{align}
where $\tau_j = \frac{j}{N} T$. 
We similarly discretize the integral of the free action as
\begin{align}
    \int_{0}^T d\tau L(z)/\hbar = \frac{T}{N} \sum_{j=0}^{N} L(z(\tau_i))/\hbar = \frac{1}{\hbar} \left( m^2 T +  \sum_{j=0}^{N-1} \frac{N}{4T} (z(\tau_{j+1})-z(\tau_j))^2 \right),
\end{align}
where the $j=0$ and $N$ terms are fixed by boundary conditions $z(\tau_0)=z(0)=x$ and $z(\tau_N)=z(T)=x'$.  
We can evaluate the resulting $D(N-1)$-dimensional integral by performing first the $z(\tau_1)$ integral, then $z(\tau_2)$, etc, with the $z(\tau_j)$ integral yielding a factor of $\left(\frac{j}{j+1} \frac{4\pi T}{N\hbar}\right)^{D/2}$. Combining these results we find 
\begin{align}
    K(x,x',T) & \equiv \int\limits_{\substack{z(0)=x\\z(T)=x'}} (Dz)_N e^{-\int_{0}^T d\tau L(z)/\hbar} \\
    &= {\cal N}_{N} N^{-D/2} \left(\frac{4\pi T}{N\hbar}\right)^{(N-1)\,D/2} e^{-(x-x')^2/(4T)- m^2T}.
\end{align}
To fix the normalization ${\cal N}_N$ for general mass $m$, it is easiest to evaluate the Fourier transform of the two-point function obtained from $K$,
\begin{align}
D(p) &= \int d^4 x e^{-ip\cdot (x-x')} D(x,x') \\
& = \int d^4 x \int_0^\infty dT e^{-ip\cdot (x-x')} K(x,x',T)\\
&= \int_0^\infty dT e^{-(p^2 + m^2) T}{\cal N}_{N} N^{-D/2} \left(\frac{4\pi T}{N\hbar}\right)^{(N-1) D/2}. 
\end{align}
To reproduce the correct two-point function, we must demand
\begin{align}
&{\cal N}_{N} N^{-D/2} \left(\frac{4\pi T}{N\hbar}\right)^{(N-1) D/2} = \left(\frac{4\pi T}{\hbar}\right)^{-D/2},\\
&\Rightarrow {\cal N}_{N} = \left(\frac{4\pi T}{N\hbar}\right)^{-N\,D/2}.
\end{align}

A final aside: It is also natural, and standard in derivations of the non-relativistic path integral, to derive the discretized path integral from $N-1$ insertions of the identity in a Hamiltonian formulation of quantum mechanics. In this approach, $N-1$ spatial integrals and $N$ momentum integrals are introduced.  The $N$ momentum integrals are Gaussian and the evaluation of each yields precisely a  normalization factor of the form ${\cal N}_N$.  

\subsection{Resolution of the Identity and Extending the Path Integral}\label{sapp:identityForExtending}
As follows from the discretization discussion above, the free Euclidean path integral in $D$ space-time dimensions reduces to
\begin{align}
  \int\limits_{\substack{z(0)=x\\z(T)=x'}} Dz e^{-\int d\tau \left(\dot z^2/4 + m^2\right)} = e^{-(x-x')^2/{4T}} (4\pi T)^{-D/2} e^{-m^2 T}. \label{freePathResult}
\end{align}
Upon integrating over all positive $T$ as in \eqref{eq:startingPI}, this recovers the standard two-point function, e.g. for $m=0$ and $D=4$, $\delta((x-x')^2)$.

However, we can instead integrate  \eqref{freePathResult} over all $x'$ (or instead over all $x$) without integrating over $T$. This is a simple Gaussian integral with result $e^{-m^2T}$.  
Thus, for $m=0$, we see that 
\begin{align}
\int d^4 x \int\limits_{\substack{z(0)=x\\z(T)=x'}} Dz e^{-\int d\tau \dot z^2/4} = 1. \label{eq:resolutionOfIdentity}
\end{align}
This is precisely the resolution of the identity claimed above \eqref{eq:matterVtx}, up to a shift in the worldline parametrization.

Now consider the Fourier transform of the first position coordinate on the two-point function \eqref{eq:startingPI}.  The Fourier transform introduces a factor
\begin{align}
\int d^4 x e^{-ip\cdot x} = \int d^4 x e^{-ip\cdot z(0)}.
\end{align}
Multiplying this factor, the original path integral, and the resolution of unity \eqref{eq:resolutionOfIdentity} but with $\tau$ running from $\tau = -T'$ to $\tau = 0$ yields (focusing on $m=0$ for simplicity)
\begin{align}
&\int d^4 x_* \int\limits_{\substack{z(-T')=x_*\\z(0)=x}} Dz 
\int d^4 x \int\limits_{\substack{z(0)=x\\z(T)=x'}} Dz 
e^{-ip\cdot z(0)} e^{-\int_{-T'}^0 d\tau \dot z^2/4 - \int_0^T d\tau \dot z^2/4} \\ 
& = \int d^4 x_* \int\limits_{\substack{z(-T')=x_*\\z(T)=x'}} Dz
V_{in}(0) e^{-\int_{-T'}^{T} d\tau \dot z^2/4}.
\end{align}
Note that, because ${\cal N}_N$ in the discretization takes the form $(4\pi \Delta \tau)^{- N\,D/2}$, if we divide the $T$ and $T'$ integrals into equal step sizes $\Delta \tau = T/N = T'/N'$ then the normalization factors in the above are also consistent: both the upper and lower expressions have a normalization $(4\pi \Delta \tau)^{-(N+N')\,D/2}$. 
Inserting a second resolution of unity for a path between $x'$ at $\tau=T$ and $x'_*$ at $\tau=T+T'$, as well as Fourier transform from  $x'$ to $p'$, yields an expression of the form \eqref{eq:path_M_QED} where Fourier phases look like vertex operators inserted in an extended path integral.  In \eqref{eq:path_M_QED} and subsequent expressions, we have dropped the boundary conditions on the $z$ path integral rather than introducing explicit limits $z(-T') = x_*$ that are then integrated over. 

\subsection{Boundary Conditions, Green's Functions, the Zero Mode, and the Momentum Conserving $\delta$-Function}
\label{sapp:BCZeroModeDelta}
We have seen that the extended path integral naturally has unconstrained endpoints. This has an important implication: the path integral includes an integration over families of paths that differ only by a $\tau$-independent translation, as well as over paths that differ by a translation that is linear in $\tau$ (i.e. a net velocity).  These motions are both annihilated by the kinetic operator $\partial_\tau^2$, i.e.~they are zero-modes.  The careful treatment of the overall translation zero-modes yields the momentum-conserving $\delta$-function support of the amplitude as noted in \eqref{eq:PdeltaFunction}. The effect of the linear-in-$\tau$ zero-mode simply cancels off the $(4\pi T)^{-D/2}$ normalization of the path integral. The derivation of this result is also coupled to the choice of Green's function, and so we briefly comment on all of these issues together.

For simplicity, we translate the extended worldline parameter $\tau$ so that the endpoints of the path integral are $\tau=0$ and $T>0$.
The path integral we wish to evaluate takes the form 
\begin{align}
\int Dz e^{-\int_{0}^T d\tau \left(\dot z(\tau)^2/4  + z(\tau) j(\tau) \right) }
\end{align}
with the endpoints unconstrained.  Up to boundary terms, we can solve this by ``completing the square''.  This logic motivates introducing a shifted worldline variable 
\begin{align}
w(\tau) \equiv z(\tau) - \int_0^{T} d\tau' G(\tau,\tau') j(\tau'),
\end{align}
where $G$ is any symmetric operator $G(\tau,\tau')=G(\tau',\tau)$ satisfying the (massless) Green's function condition
\begin{align}
\frac{1}{2} \partial_\tau^2 G(\tau,\tau') = \delta(\tau-\tau').
\label{eq:GreensFnCondition}
\end{align}
We will elaborate on the space of solutions below.  But first, let us see what happens when we rewrite the action and worldline path integral in terms of $w$ --- now keeping track of boundary terms. 

Since at each time $\tau$, $w$ is simply a ($\tau$-dependent) shift of $z$, the measure $Dw = D z$.
By substituting for $z$ in terms of $w$, integrating by parts, and using the defining property \eqref{eq:GreensFnCondition} of the Green's function $G$, we find that 
\begin{align}
   & \int d\tau \left(\dot z(\tau)^2/4  + z(\tau) j(\tau) \right) \\ &= 
    \int d\tau \tfrac{1}{4} \dot w(\tau)^2 + 
    \int d\tau d\tau' \tfrac{1}{2} j(\tau) G(\tau,\tau') j(\tau') + 
[\tfrac{1}{2} \int d\tau' w(\tau) \partial_\tau G(\tau,\tau') j(\tau')] \\
& \qquad + [\tfrac{1}{4} \int d\tau' d\tau'' G(\tau,\tau'') \partial_\tau G(\tau,\tau') j(\tau') j(\tau'')],
\label{eq:xjIntByPartsResult}
\end{align}
where we use $[f(\tau,\ldots)] \equiv f(\tau=T,\ldots) - f(\tau=0,\ldots)$ to denote boundary terms from $\tau$-integration by parts.  
As expected, the action depends on $w(\tau)$ only through (linear) boundary terms and the free action $\dot{w}(\tau)^2/4$.  Thus, we can solve the path integral using \eqref{freePathResult}.  Writing the integrals over the boundary values $w_- = w(0)$ and $w_+ = w(T)$ explicitly, we have
\begin{align}
\int Dz &e^{-\int d\tau \left(\dot z(\tau)^2/4  + z(\tau) j(\tau) \right) } \\ 
&= \int d^4w_- d^4 w_+
\int\limits_{\substack{w(0)=w_-\\w(T)=w_+}} Dw 
e^{-\int d\tau \dot w(\tau)^2/4 -
    \int d\tau d\tau' \tfrac{1}{2} j(\tau) G(\tau,\tau') j(\tau') }\\
    &\qquad\qquad\qquad\qquad\qquad\qquad e^{-
\left[\tfrac{1}{2} \int d\tau' \left(w(\tau) + \tfrac{1}{2} \int d\tau'' G(\tau,\tau'') j(\tau'')\right) \partial_\tau G(\tau,\tau') j(\tau')\right] }
\\
&= e^{-\tfrac{1}{2} \int d\tau d\tau' j(\tau) G(\tau,\tau') j(\tau')} (4\pi T)^{-2} \\
&\quad \quad\int d^4w_- d^4w_+
e^{-(w_--w_+)^2/{4T} 
-[((w_+(\tau) + \tfrac{1}{2} \int d\tau'' G(\tau,\tau'') j(\tau'')) \int d\tau' \partial_\tau G(\tau,\tau') j(\tau')]}\\
&=e^{-\tfrac{1}{2} \int d\tau d\tau' j(\tau) G(\tau,\tau') j(\tau')} (4\pi T)^{-2} 
\int d^4w_- d^4w_+ e^{-(w_--w_+)^2/{4T}
-[((w(\tau) + I_0(\tau)) I_1(\tau)]} \label{eq:finalWPI}
\end{align}
where we have introduced the shorthand
\begin{align}
I_n(\tau)\equiv \tfrac{1}{2} \int d\tau' (\partial_\tau)^n G(\tau,\tau') j(\tau')
\end{align}
for  combinations that appear in the boundary terms, which importantly are \emph{$z_\pm$-independent} and \emph{linear in $j$}. 

Let us return now to the solutions of the Green's function condition \eqref{eq:GreensFnCondition} that appear in the expression above. 
Since this is a second-order equation, the general solution has two free parameters and can be written as
\begin{align}
    G(\tau,\tau') = |\tau-\tau'| - a (\tau+\tau') + 2 b \frac{\tau\tau'}{T}.
\end{align}
The choice $a=b=0$, used in the main text, is the unique translation-invariant Green's function which we will denote here as $\bar G(\tau-\bar\tau)$. Another common choice in the literature is $a=b=1$, which enforces Dirichlet boundary conditions at both $\tau=0$ and $T$.

Since we are working with a path integral over unconstrained $x$, we need not impose any specific boundary condition and can choose a Green's function at our convenience.  Different choices of $a$ and $b$ alter both the $\int d\tau d\tau' j(\tau) G(\tau,\tau') j(\tau')$ term and the integeration-by-parts boundary terms, but in such a way that all $a$ and $b$-dependence drops out from the final result (as it must, since the result for \emph{any} $a$ and $b$ must equal the original $z$ path integral).  In terms of $\bar G$, and rewriting the endpoint integrals in terms of CM and relative coordinates $W=(w_- + w_+)/2$ and $w_r = w_+ - w_-$, the result is
\begin{align}
\int d^4 W d^4 w_r (4\pi T)^{-2}  e^{-\frac{1}{2} \int d\tau d\tau' j(\tau) \bar G(\tau-\tau') j(\tau') - \int {d\tau} (i W P  + \frac{1}{4T} \left(w_r - i T (P + 2 Q))\right)^2 + T P^2/4}
\end{align}
where
\begin{align}
    P\equiv -i \int d\tau j(\tau) \quad Q \equiv -i \int d\tau j(\tau) \tau/T.\label{eq:PQdef}
\end{align}
Upon performing the Gaussian $w_r$ integral and exponential $W$ integral, we obtain 
\begin{align}
    (2\pi)^4 \delta^{(4)}(P) e^{-\frac{1}{2} \int d\tau d\tau' j(\tau) \bar G(\tau-\tau') j(\tau') }
\end{align}
as claimed in Eq.~\eqref{eq:PdeltaFunction}.

Now, consider inserting the explicit form of the current \eqref{eq:jdefinition} in \eqref{eq:PQdef} we find $P=p_0+p_3+k_1+k_2$ --- the sum of all outgoing momenta --- from $\bar j$ and no contribution from $\hat j$.  This generalizes to higher-point amplitudes: $\delta^{(4)}(P)$ is always the four-momentum-conservation $\delta$-function, which naturally arises from integrating over all possible space-time translations of a given worldline path. 

Besides elucidating the origin of the momentum-conserving $\delta$-function, the above discussion also motivates our use of the $\tau$-translation-invariant Green's function $\bar G$ as the simplest choice.  Had we used a general Green's function, then the first term in \eqref{eq:finalWPI} would have received a correction $Q (bQ -a P)$, which however would be canceled by boundary terms in the final result.

\section{Useful Identities Involving Products of Distributions}
\label{app:distributions}

To evaluate path integrals, we will frequently need to integrate over distributions and discontinuous functions, such as $\delta(x), \theta(x)$ and so on.  This must be done with some care.  In general, the path integral as presented here generates products of distributions. As is well known, these cannot be unambiguously defined whenever singularities overlap with one another or with discontinuities.  Below, we motivate and describe the conventions we adopt.  These prescriptions yield reasonable physical amplitudes (finite, little-group-covariant, and reproducing standard results in the $\rho\to 0$ limit) but we do not have a rigorous first-principles derivation of these rules --- particularly the prescription of dropping overlapping products of $\delta$-functions. One may speculate that these ambiguous terms could be more cleanly removed, without changing the result, by extending the worldline path integral to include either additional physical modes of the matter, motivated by the finite extent of the CSP currents as discussed in \cite{Schuster:2023xqa}, or ghost modes, as occurs in the worldline computation of amplitudes in general relativity.

\subsubsection*{Overlapping Singularities}
Overlapping singularities, such as $\delta(\tau)^2$, are not well-defined as distributions. To resolve this, we adopt what is sometimes referred to as the ``Veltman prescription'' in the literature, where integrals of products of delta-functions are taken to vanish at coincident points. This is a natural consequence of requiring distributions to give continuous results under infinitesimal shifts of their arguments. In general, this prescription can be motivated by the requirement that, if an $\epsilon$-shift in the argument of a $\delta$-function would lead to a vanishing product for arbitrarily small $\epsilon$, then the product must vanish. 
Thus, we take
\begin{align}
    \delta(t)^2 \to \delta(t) \delta(t\pm \epsilon) = 0 \\ 
    \delta(0) \to \delta(\epsilon) = 0. 
\end{align}
This prescription yields a self-consistent way of defining products of singular distributions, such as $\delta(x)$. It is also the unique prescription that is independent of any regulator used for defining $\delta(x)$ in integrals of smooth functions.   

\subsubsection*{Singularities Overlapping Discontinuities}
Another ambiguity arises when delta-functions are multiplied by functions with a discontinuity overlapping the $\delta$-function. However, integrals of such expressions can be evaluated by using the Leibnitz rule. For example, since $\frac{d}{dx} \sign(x) = 2 \delta(x)$,
\begin{align}
 \int_{0}^\infty dx \delta(x) \sign(x) &= \int_{0}^\infty dx \frac{d}{dx}{\sign^2(x)/4}\\
& = \left[\sign^2(x)/4\right]^{\infty}_0 = \frac{1}{4}
\end{align}
where we have taken $\sign(0)=0$ by parity. This generalizes readily, but with a non-trivial factor, to higher powers of $\sign(x)$: 
\begin{align}
& \int_{0}^\infty dx \delta(x) \sign(x)^p = \int_{0}^\infty dx \frac{d}{dx}\left(\frac{\sign(x)^{p+1}}{2(p+1)}\right) = \frac{1}{2(p+1)}.
\end{align}
Therefore, we see that in integrals involving $\delta(x)$ we cannot simply replace $\sign(x)^{2n}\to 1$ or $\sign(x)^{2n+1}\to \sign(x)$. 

We will also encounter integrals of the form
\begin{equation}
    I_p \equiv \int_0^\infty dx \delta(x) e^{-a \sign(x) -b x} (1-\sign(x))^p
\end{equation}
where $b>0$.
We can solve for $I_0$ by considering 
\begin{align}
 \int_0^\infty & dx  \frac{d}{dx}\left\{  e^{-a \sign(x) -b x} \right\} \\
    & = \int_0^\infty dx  (-2a\delta(x) -b)  e^{-a \sign(x) -b x}  \\
    & = -2a I_0 -b \int_0^\infty dx e^{-a -b x} = -2a I_0 - e^{-a},
\end{align}
where in the last line we have used $\sign(x)=+1$ on the domain of integration, valid since the integrand is not singular at zero. On the other hand, the same expression can be evaluated using the Leibnitz rule to obtain
\begin{align}
\left[e^{-a \sign(x) -b x}\right]^{\infty}_{x=0} = -1, \label{eqref:leibnitz}
\end{align}
with the contribution from the upper endpoint vanishing for $b>0$ and the lower endpoint giving -1 since $\sign(x)=0$.
Identifying the two results above and solving for $I_0$ gives
\begin{align}
    I_0 = \frac{1}{2a} (1-e^{-a}) = \frac{1}{2} \int_0^1 dx e^{-ax}.
\end{align}
Furthermore, since $I_{p} = (1-\frac{d}{da}) I_{p-1}$ for $p>0$, we find
\begin{align}
    I_p = \frac{1}{2} \int_0^1 dx (1-x)^p e^{-ax}.
\end{align}
This is a particularly useful identity that we make use of in our path integral calculations. Similar logic gives us 
\begin{align}
    J_0 & \equiv \int_{-\infty}^{\infty} dx 2\delta(x) e^{-a \sign(x) -b x - c x \sign\,x} 
    & = \frac{1}{a} \left(e^a - e^{-a}\right) = \int_{-1}^1 dx e^{ax},
\end{align}
with $b+c>0$ and $c-b>0$ required for the removal of boundary terms as in \eqref{eqref:leibnitz}.  This is the form that we will encounter in our parametrization of the path integral. Had we broken up \eqref{eq:2emissionPathIntegral} into a sum over orderings of $t_1$ and $t_2$, which  is sometimes done in the literature and more closely resembles the Feynman diagram expansion, we would instead encounter integrals of the form $I_0$. 

\section{Yukawa-Like Vertex Operator Compton Scattering}
\label{sec:scalarLike}

The auxiliary form \eqref{eq:scalarAmplitudeSimplifiedBeforeMuIntegrals} is equally applicable to massless Yukawa-like CSP-matter couplings if we take $w=0$ rather than $w=1$ in the auxiliary integrals.  In this case \eqref{eq:scalarAmplitudeSimplifiedBeforeMuIntegrals}  yields
\begin{align}
M_{LSZ} &= 2 \int_{-1}^{1} dx \left( \frac{d}{dx} \frac{1}{k_2\cdot P_2(x)} - \frac{d}{d P_1(x)}\cdot \frac{d}{dP_2(x)} \right)  e^{-i\rho\frac{\eta_1 \cdot P_1(x)}{k_1\cdot P_1(x) } -i\rho\frac{\eta_2 \cdot P_2(x)}{k_2\cdot P_2(x) }} \label{scalarMFunctionUnsimplified} \\ 
& = 2 \left[\frac{1}{k_2\cdot P_2(x)} e^{-i\rho\frac{\eta_1 \cdot P_1(x)}{k_1\cdot P_1(x) } -i\rho\frac{\eta_2 \cdot P_2(x)}{k_2\cdot P_2(x) }} \right]_{x=-1}^{x=+1} \\
&\qquad + 2 \rho^2 \int_{-1}^1 dx \bigg\{\left( \frac{\eta_1}{ k_1 \cdot P_1(x)} - \frac{\eta_1\cdot P_1(x)\,k_1 }{( k_1 \cdot P_1(x))^2}\right) 
\left( \frac{\eta_2}{ k_2 \cdot P_2(x)} - \frac{\eta_2\cdot P_2(x)\,k_2 }{( k_2 \cdot P_2(x))^2}\right)\\
& \qquad \qquad \qquad e^{-i\rho\frac{\eta_1 \cdot P_1(x)}{k_1\cdot P_1(x) } -i\rho\frac{\eta_2 \cdot P_2(x)}{k_2\cdot P_2(x)}} \bigg\}.\label{eq:scalarMFunctionPartiallySimplified}
\end{align}

The first term evaluates to
\begin{align}
\frac{1}{s} e^{-i\rho \frac{\eta_1 \cdot (2 p_0 + k_1)}{k_1\cdot (2 p_0 + k_1)}} e^{-i\rho \frac{\eta_2 \cdot (2 p_3 + k_2)}{k_2\cdot (2 p_3 + k_2)}} + \frac{1}{u} e^{-i\rho \frac{\eta_1 \cdot (2 p_3 + k_1)}{k_1\cdot (2 p_3 + k_1)}} e^{-i\rho \frac{\eta_2 \cdot (2 p_0 + k_2)}{k_2\cdot (2 p_0 + k_2)}},\label{eq:boundaryTermScalarMFunction}
\end{align}
which, incidentally, matches what one might have obtained from Feynman rules with a CSP ``vertex'' (for all momenta outgoing) $V({k,\eta},p,p') = e^{-i\rho \frac{\eta \cdot (p-p')}{k\cdot (p-p')}}$ (though we note that this vertex ansatz on its own is ill-behaved when $p$ and $p'$ are both taken on-shell, since $k\cdot(p-p') = p^2 - {p'}^2 = 0$ in this case). 

The second term, which starts at ${\cal O}(\rho^2)$, can be evaluated exactly in terms of elementary functions and is meromorphic (in contrast with the vector case which contains an exponential integral), though the final expression is long and not particularly enlightening.  To reach this closed form, it is useful to change variables from $x$ to $v=1/r(x)$, where $r = \equiv k_2 \cdot P_2(x) = -k_1 \cdot P_1(x)$ as before.  Because each contribution to this term involves at least two inverse powers of $r$, the integrand after this change of variables contains only non-negative powers of $v$ multiplying the exponential
$e^{i\rho v(\eta_1 \cdot P_1(x)-\eta_2 \cdot P_2(x))}$, the phase of which is linear in $v$.  

\bibliography{CSPAmplitudes}

\end{document}